# Statistical Quadrature Evolution by Inference for Continuous-Variable Quantum Key Distribution

## Laszlo Gyongyosi

[1] Quantum Technologies Laboratory

Department of Networked Systems and Services

Budapest University of Technology and Economics

2 Magyar tudosok krt., Budapest, H-1117 Hungary

[2] MTA-BME Information Systems Research Group

Hungarian Academy of Sciences

7 Nador st., Budapest, H-1051 Hungary

gyongyosi@hit.bme.hu

**Abstract**

We define the statistical quadrature evolution (QE) method for multicarrier continuous-variable quantum key distribution (CVQKD). A multicarrier CVQKD protocol uses Gaussian subcarrier quantum continuous variables (CVs) for information transmission. The QE scheme utilizes the theory of mathematical statistics and statistical information processing. The QE model is based on the Gaussian quadrature inference (GQI) framework to provide a minimal error estimate of the CV state quadratures. The QE block evaluates a unique and stable estimation of the non-observable continuous input from the measurement results and through the statistical inference method yielded from the GQI framework. The QE method minimizes the overall expected error by an estimator function and provides a viable, easily implementable, and computationally efficient way to maximize the extractable information from the observed data. The QE framework can be established in an arbitrary CVQKD protocol and measurement setting and is implementable by standard low-complexity functions, which is particularly convenient for experimental CVQKD.

# 1 Introduction

Continuous-variable quantum key distribution (CVQKD) provides a method to realize unconditional secure communication over standard, currently established telecommunication networks [10–22]. A significant attribute of CVQKD is that, in contrast to DV (discrete variable) QKD, it does not require single-photon sources and detectors and can be implemented by standard devices [1], [9–26], [30–37]. In a CVQKD setting, the information is carried by a continuous-variable quantum state that is defined in the phase space via the position and momentum quadratures. In a practical CVQKD implementation, the CV quantum states have a Gaussian random distribution, and the quantum channel between the sender (Alice) and receiver (Bob) is also Gaussian, because the presence of an eavesdropper (Eve) adds a white Gaussian noise into the transmission [19-21].

The CVQKD protocols have several attractive properties. However, the relevant performance attributes, such as secret key rates and transmission distances, still require significant improvements. For this purpose, the multicarrier CVQKD has been recently introduced through the adaptive quadrature division modulation (AMQD) scheme [2]. The multicarrier CVQKD injects several additional degrees of freedom into the transmission, which are not available for a standard, single-carrier CVQKD setting. In particular, these extra benefits and resources allow the realization of higher secret key rates and a higher amount of tolerable losses with unconditional security. These innovations opened a door to the establishment of several new phenomena for CVQKD that are unrealizable in standard CVQKD, such as single layer transmission [4], enhanced security thresholds [5], multidimensional manifold extraction [6], characterization of the subcarrier domain [7], adaptive quadrature detection and sub-channel estimation techniques [8], and an extensive utilization of distribution statistics and random matrix formalism [9]. The benefits of multicarrier CVQKD have also been proposed for multiple access multicarrier CVQKD via the AMQD-MQA (multiuser quadrature allocation) [3]. A statistical information processing model of multicarrier CVQKD and the Gaussian quadrature inference (GQI) methods were proposed in [10]. The GQI provides an optimal statistical estimation of the non-observable Gaussian subcarrier input quadratures from the observed noisy Gaussian subcarriers using the theory of statistical information processing [25-28].

In this work, we propose the statistical *quadrature evolution* (QE) method for multicarrier CVQKD. The QE extends the results of the GQI to evolve a minimal error estimate of the continuous input regime from noisy discrete variables and estimates. The QE method utilizes the theory of mathematical statistics and the fundamentals of statistical information processing. While the GQI operates on the measured noisy subcarrier CVs to estimate the continuous spectrum of the non-observable input subcarriers, the QE scheme functionally builds on the GQI output and on the measured discrete noisy subcarrier variables to evolve the continuous regime of the non-observable single-carrier inputs. The QE method achieves a theoretically minimized magnitude error to evolve the non-observable single-carrier continuous variable Gaussian quadratures from the observed discrete variables. The theoretical minimum error of QE is function-ally provided by the GQI method [10] and by an optimal estimator function applied to the measurement results.

The QE block evaluates a unique estimation of the non-observable continuous input regime via the noisy subcarrier variables yielded from the measurements and by the subcarrier estimations yielded from the GQI. Precisely, the QE model utilizes a corresponding estimator function that provides an optimal least-square formulation (optimal least-squares estimator). In fact, the estimator function is a linear operator whose coefficients depend on the GQI output. The QE framework also solves the problem of the optimal least-square estimate being a non-linear function of the observed data (e.g., discrete subcarrier variables and GQI output). Particularly, it is explained by the mathematical fact that the optimal least-square estimator is a linear function of the observed data if both the non-observable input (Gaussian CV states) and the observed data (measured CV states) are jointly Gaussian [28-29], which is exactly the case in a CVQKD setting.

The QE framework provides a viable, easily implementable, and computationally efficient way to maximize the extractable information from the observed data. Using the fundamental theory that stands behind the GQI and QE blocks, the proposed QE output provided by the least-square formulation is always a unique and stable solution, which minimizes the overall expected estimation error for each individual component of the input vector. Specifically, by exploiting the statistical framework of multicarrier CVQKD, we prove that QE in a multicarrier CVQKD setting achieves a vanishing error as the number of subcarriers dedicated to a given user increases. We derive the corresponding expected error and expected variance using the statistical model of multicarrier CVQKD and demonstrate the results through numerical evidence. The error and error variances and the corresponding covariances are derived via computationally efficient, easily implementable functions.

This paper is organized as follows. In Section 2, preliminary findings are summarized. Section 3 discusses the QE method for multiple access multicarrier CVQKD, and derives the achievable statistical secret key rates. Finally, Section 4 concludes the results. A numerical evidence is included in the Supplemental Information.

## 2 Preliminaries

In Section 2, the notations and basic terms are summarized. For further information, see the detailed descriptions of [2–10].

### 2.1 Multicarrier CVQKD

The following description assumes a single user, and the use of $n$ Gaussian sub-channels $\mathcal{N}_i$ for the transmission of the subcarriers, from which only $l$ sub-channels will carry valuable information.

In the single-carrier modulation scheme, the $j$-th input single-carrier state $\left|\varphi_j\right\rangle = \left|x_j + \mathrm{i}p_j\right\rangle$ is a

Gaussian state in the phase space $\mathcal{S}$, with i.i.d. Gaussian random position and momentum quadratures

$$x_j \in \mathbb{N}\left(0, \sigma_{\omega_0}^2\right), \ p_j \in \mathbb{N}\left(0, \sigma_{\omega_0}^2\right), \tag{1}$$

where $\sigma_{\omega_0}^2$ is the modulation variance of the quadratures. In the multicarrier scenario, the information is carried by Gaussian subcarrier CVs, $\left|\phi_i\right\rangle = \left|x_i + \mathrm{i}p_i\right\rangle$, via quadratures

$$x_i \in \mathbb{N}\left(0, \sigma_{\omega}^2\right), \ p_i \in \mathbb{N}\left(0, \sigma_{\omega}^2\right), \tag{2}$$

where $\sigma_{\omega}^2$ is the modulation variance of the subcarrier quadratures, which are transmitted through a noisy Gaussian sub-channel $\mathcal{N}_i$. Precisely, each $\mathcal{N}_i$ Gaussian sub-channel is dedicated for the transmission of one Gaussian subcarrier CV from the $n$ subcarrier CVs. (*Note*: index $i$ refers to a subcarrier CV, index $j$ to a single-carrier CV, respectively.)

The single-carrier CV state $\left|\varphi_j\right\rangle$ in $\mathcal{S}$ can be modeled as a zero-mean, circular symmetric complex Gaussian random variable $z_j \in \mathcal{CN}\left(0, \sigma_{\omega_{z_j}}^2\right)$, with a variance

$$\sigma_{\omega_{z_j}}^2 = \mathbb{E}\left[\left|z_j\right|^2\right] = 2\sigma_{\omega_0}^2, \tag{3}$$

and with i.i.d. real and imaginary zero-mean Gaussian random components

$$\mathrm{Re}\left(z_j\right) \in \mathbb{N}\left(0, \sigma_{\omega_0}^2\right), \ \mathrm{Im}\left(z_j\right) \in \mathbb{N}\left(0, \sigma_{\omega_0}^2\right). \tag{4}$$

In the multicarrier CVQKD scenario, let $n$ be the number of Alice's input single-carrier Gaussian states. Precisely, the $n$ input coherent states are modeled by an $n$-dimensional, zero-mean, circular symmetric complex random Gaussian vector

$$\mathbf{z} = \mathbf{x} + \mathrm{i}\mathbf{p} = \left(z_0, \ldots, z_{n-1}\right)^T \in \mathcal{CN}\left(0, \mathbf{K_z}\right), \tag{5}$$

where each $z_j$ is a zero-mean, circular symmetric complex Gaussian random variable

$$z_j \in \mathcal{CN}\left(0, \sigma_{\omega_{z_j}}^2\right), \ z_j = x_j + \mathrm{i}p_j. \tag{6}$$

In the first step of AMQD, Alice applies the inverse FFT (fast Fourier transform) operation to vector $\mathbf{z}$ (see (5)), which results in an $n$-dimensional zero-mean, circular symmetric complex Gaussian random vector $\mathbf{d}$, $\mathbf{d} \in \mathcal{CN}\left(0, \mathbf{K_d}\right)$, $\mathbf{d} = \left(d_0, \ldots, d_{n-1}\right)^T$, precisely as

$$\mathbf{d} = F^{-1}\left(\mathbf{z}\right) = e^{\frac{\mathbf{d}^T \mathbf{A}\mathbf{A}^T \mathbf{d}}{2}} = e^{\frac{\sigma_{\omega_0}^2\left(d_0^2 + \ldots + d_{n-1}^2\right)}{2}}, \tag{7}$$

where

$$d_i = x_{d_i} + \mathrm{i}p_{d_i}, \ d_i \in \mathcal{CN}\left(0, \sigma_{d_i}^2\right), \tag{8}$$

where $\sigma_{\omega_{d_i}}^2 = \mathbb{E}\left[\left|d_i\right|^2\right] = 2\sigma_{\omega}^2$, thus the position and momentum quadratures of $\left|\phi_i\right\rangle$ are i.i.d. Gaussian random variables with a constant variance $\sigma_{\omega}^2$ for all $\mathcal{N}_i, i = 0, \ldots, l-1$ sub-channels:

$$\operatorname{Re}\left(d_i\right) = x_{d_i} \in \mathbb{N}\left(0, \sigma_\omega^2\right), \ \operatorname{Im}\left(d_i\right) = p_{d_i} \in \mathbb{N}\left(0, \sigma_\omega^2\right), \tag{9}$$

where $\mathbf{K_d} = \mathbb{E}\left[\mathbf{dd}^\dagger\right]$, $\mathbb{E}\left[\mathbf{d}\right] = \mathbb{E}\left[e^{\mathrm{i}\gamma}\mathbf{d}\right] = \mathbb{E}e^{\mathrm{i}\gamma}\left[\mathbf{d}\right]$, $\mathbb{E}\left[\mathbf{dd}^T\right] = \mathbb{E}\left[e^{\mathrm{i}\gamma}\mathbf{d}\left(e^{\mathrm{i}\gamma}\mathbf{d}\right)^T\right] = \mathbb{E}e^{\mathrm{i}2\gamma}\left[\mathbf{dd}^T\right]$ for any $\gamma \in \left[0, 2\pi\right]$.

The $\mathbf{T}\left(\mathcal{N}\right)$ transmittance vector of $\mathcal{N}$ in the multicarrier transmission is

$$\mathbf{T}\left(\mathcal{N}\right) = \left[T_0\left(\mathcal{N}_0\right), ..., T_{n-1}\left(\mathcal{N}_{n-1}\right)\right]^T \in \mathcal{C}^n, \tag{10}$$

where

$$T_i\left(\mathcal{N}_i\right) = \operatorname{Re}\left(T_i\left(\mathcal{N}_i\right)\right) + \mathrm{i}\operatorname{Im}\left(T_i\left(\mathcal{N}_i\right)\right) \in \mathcal{C}, \tag{11}$$

is a complex variable, which quantifies the position and momentum quadrature transmission (i.e., gain) of the $i$-th Gaussian sub-channel $\mathcal{N}_i$, in the phase space $\mathcal{S}$, with real and imaginary parts

$$0 \leq \operatorname{Re} T_i\left(\mathcal{N}_i\right) \leq 1/\sqrt{2}, \ \text{and} \ 0 \leq \operatorname{Im} T_i\left(\mathcal{N}_i\right) \leq 1/\sqrt{2}. \tag{12}$$

Particularly, the $T_i\left(\mathcal{N}_i\right)$ variable has the squared magnitude of

$$\left|T_i\left(\mathcal{N}_i\right)\right|^2 = \operatorname{Re} T_i\left(\mathcal{N}_i\right)^2 + \operatorname{Im} T_i\left(\mathcal{N}_i\right)^2 \in \mathbb{R}, \tag{13}$$

where

$$\operatorname{Re} T_i\left(\mathcal{N}_i\right) = \operatorname{Im} T_i\left(\mathcal{N}_i\right). \tag{14}$$

The Fourier-transformed transmittance of the $i$-th sub-channel $\mathcal{N}_i$ (resulted from CVQFT operation at Bob) is denoted by

$$\left|F\left(T_i\left(\mathcal{N}_i\right)\right)\right|^2. \tag{15}$$

The $n$-dimensional zero-mean, circular symmetric complex Gaussian noise vector $\Delta \in \mathcal{CN}\left(0, \sigma_\Delta^2\right)_n$, of the quantum channel $\mathcal{N}$, is evaluated as

$$\Delta = \left(\Delta_0, ..., \Delta_{n-1}\right)^T \in \mathcal{CN}\left(0, \mathbf{K}_\Delta\right), \tag{16}$$

where

$$\mathbf{K}_\Delta = \mathbb{E}\left[\Delta\Delta^\dagger\right], \tag{17}$$

with independent, zero-mean Gaussian random components

$$\Delta_{x_i} \in \mathbb{N}\left(0, \sigma_{\mathcal{N}_i}^2\right), \ \text{and} \ \Delta_{p_i} \in \mathbb{N}\left(0, \sigma_{\mathcal{N}_i}^2\right), \tag{18}$$

with variance $\sigma_{\mathcal{N}_i}^2$, for each $\Delta_i$ of a Gaussian sub-channel $\mathcal{N}_i$, which identifies the Gaussian noise of the $i$-th sub-channel $\mathcal{N}_i$ on the quadrature components $x_i, p_i$ in the phase space $\mathcal{S}$. Thus $F\left(\Delta\right) \in \mathcal{CN}\left(0, \sigma_{\Delta_i}^2\right)$, where

$$\sigma_{\Delta_i}^2 = 2\sigma_{\mathcal{N}_i}^2. \tag{19}$$

The CVQFT-transformed noise vector can be rewritten as

$$F\left(\Delta\right) = \left(F\left(\Delta_0\right),...,F\left(\Delta_{n-1}\right)\right)^T, \tag{20}$$

with independent components $F\left(\Delta_{x_i}\right) \in \mathbb{N}\left(0, \sigma^2_{\mathcal{N}_i}\right)$ and $F\left(\Delta_{p_i}\right) \in \mathbb{N}\left(0, \sigma^2_{\mathcal{N}_i}\right)$ on the quadratures, for each $F\left(\Delta_i\right)$. Precisely, it also defines an $n$-dimensional zero-mean, circular symmetric complex Gaussian random vector $F\left(\Delta\right) \in \mathcal{CN}\left(0, \mathbf{K}_{F\left(\Delta\right)}\right)$. The complex $A_j\left(\mathcal{N}_j\right) \in \mathbb{C}$ single-carrier channel coefficient is derived from the $l$ Gaussian sub-channel coefficients as

$$A_j\left(\mathcal{N}_j\right) = \frac{1}{l} \sum_{i=0}^{l-1} F\left(T_i\left(\mathcal{N}_i\right)\right). \tag{21}$$

### 2.1.1 Multiuser Quadrature Allocation (MQA)

In a MQA multiple access multicarrier CVQKD, a given user $U_k, k = 0,...,K-1$, where $K$ is the number of total users, is characterized via $m$ subcarriers, formulating an $\mathcal{M}_{U_k}$ logical channel of $U_k$,

$$\mathcal{M}_{U_k} = \left[\mathcal{N}_{U_k,0},...,\mathcal{N}_{U_k,m-1}\right]^T, \tag{22}$$

where $\mathcal{N}_{U_k,i}$ is the $i$-th sub-channel of $\mathcal{M}_{U_k}$. For a detailed description of MQA for multicarrier CVQKD see [3].

The general model of AMQD-MQA is depicted in Fig. 1 [3], [5].

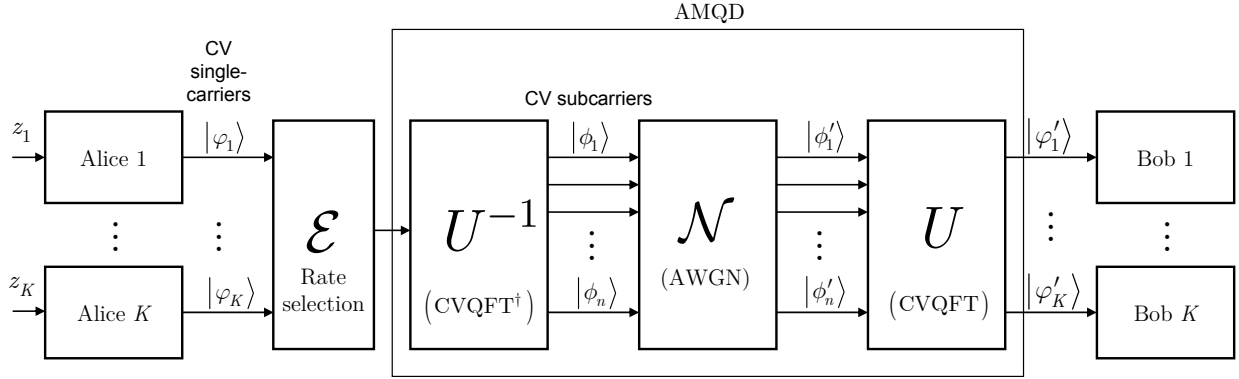

**Figure 1.** The AMQD-MQA multiple access scheme with multiple independent transmitters and multiple receivers [3]. The modulated Gaussian CV single-carriers are transformed by a unitary operation (inverse CVQFT) at the $\mathcal{E}$ encoder, which outputs the $n$ Gaussian subcarrier CVs for the transmission. The parties send the $\left|\varphi_k\right\rangle$ single-carrier Gaussian CVs with variance $\sigma^2_{\omega_{0,k}}$ to Alice. In the rate-selection phase, the encoder determines the transmit users. The data states of the transmit users are then fed into the CVQFT$^\dagger$ operation. The $\left|\phi_i\right\rangle$ Gaussian subcarrier CVs have a variance $\sigma^2_\omega$ per quadrature components. The Gaussian CVs are decoded by the CVQFT unitary operation. Each $\left|\varphi'_k\right\rangle$ is received by Bob $k$.

## 2.2 Gaussian Quadrature Inference (GQI)

In this section we summarize the basic terms of GQI, for the further details see [10].

According to the GQI framework, the $m$ $x'_{U_k,i}, i = 0,\ldots,m-1$, noisy subcarrier CVs of $U_k$, $k = 0,\ldots,K-1$, yield the $E\left(U^{-1}\left(x_{U_k,j}\right)\right)$ estimate of $U^{-1}\left(x_{U_k,j}\right)$, where $x_{U_k,j}$ is the quadrature component of $\varphi_{U_k,j}$, $\varphi_{U_k,j}$ is the $j$-th input CV of $U_k$, $\varphi_{U_k,j} = x_{U_k,j} + \mathrm{i}p_{U_k,j}$, $\left\{x_{U_k,j}, p_{U_k,j}\right\}$ are Gaussian random quadratures, as

$$E\left(U^{-1}\left(x_{U_k,j}\right)\right) = \left| 1 \Big/ -\sum_{i=0}^{m-1}\left|\mathcal{T}_i\left(e^{\mathrm{i}\theta_{\varphi_{U_k,j}}}\right)\right|^2 \tilde{\lambda}_i \right|, \tag{23}$$

where function $\mathcal{T}_i\left(e^{\mathrm{i}\theta_{\varphi_{U_k,j}}}\right)$ evaluates $T_{U_k,i}\left(\mathcal{N}_{U_k,i}\right)$ of $\mathcal{N}_{U_k,i}$, $\mathcal{N}_{U_k,i}$ is the $i$-th sub-channel of $U_k$, while $\tilde{\lambda}_i$ are optimal Lagrange multipliers.

Let

$$\varphi_{U_k,j} \in \mathbb{N}\left(0, \mathbb{E}\left\|\varphi_{U_k,j}\right\|^2 = 2\sigma_{\omega_0}^2\right), \tag{24}$$

where $x_{U_k,j} \in \mathbb{N}\left(0, \sigma_{\omega_0}^2\right)$, $p_{U_k,j} \in \mathbb{N}\left(0, \sigma_{\omega_0}^2\right)$ are Gaussian random quadratures, $\sigma_{\omega_0}^2$ is the single-carrier modulation variance [2], and let the $m$ subcarrier CVs of $U_k$ be referred via

$$\vec{\phi}_{U_k} = \left[\phi_{U_k,0} \ldots \phi_{U_k,m-1}\right]^T, \tag{25}$$

where

$$\phi_{U_k,i} = x_{U_k,i} + \mathrm{i}p_{U_k,i}, \phi_{U_k,i} \in \mathbb{N}\left(0, \mathbb{E}\left\|\phi_{U_k,i}\right\|^2 = 2\sigma_{\omega_i}^2\right), \tag{26}$$

while $x_{U_k,i} \in \mathbb{N}\left(0, \sigma_{\omega_i}^2\right)$, $p_{U_k,i} \in \mathbb{N}\left(0, \sigma_{\omega_i}^2\right)$ are the subcarrier quadratures, $\sigma_{\omega_i}^2$ is the quadrature modulation variance (chosen to be constant $\sigma_{\omega_i}^2 = \sigma_\omega^2$ for $\forall i$, see [2]), while $\mathcal{M}_{U_k}$ is the logical channel (a set of $m$ sub-channels) of $U_k$, see (22).

The output of $\mathcal{N}_{U_k,i}$ is

$$\phi'_{U_k,i} \in \mathbb{N}\left(0, 2\left(\sigma_{\omega_i}^2 + \sigma_{\mathcal{N}_{U_k,i}}^2\right)\right), \tag{27}$$

where $\sigma_{\mathcal{N}_{U_k,i}}^2$ is the noise variance of $\mathcal{N}_{U_k,i}$, and

$$\vec{\phi}'_{U_k} = \left[\phi'_{U_k,0} \ldots \phi'_{U_k,m-1}\right]^T, \tag{28}$$

where $\phi'_{U_k,i} = x'_{U_k,i} + \mathrm{i}p'_{U_k,i}$ and $x'_{U_k,i}, p'_{U_k,i}$ are noisy Gaussian random quadratures, $x'_{U_k,i} \in \mathbb{N}\left(0, \sigma_\omega^2 + \sigma_{\mathcal{N}_i}^2\right)$, $p'_{U_k,i} \in \mathbb{N}\left(0, \sigma_\omega^2 + \sigma_{\mathcal{N}_i}^2\right)$.

Let $n$ be the number of single-carriers, $n \to \infty$, and let

$$\theta_{\varphi_{U_k,j}} = \pi \Big/ \Omega, \tag{29}$$

where

$$\Omega = \sigma_{\omega_0}^2 \big/ \sigma_\omega^2 \, , \qquad (30)$$

and where $\sigma_{\omega_0}^2$, $\sigma_\omega^2$ are the single-carrier and multicarrier modulation variances, respectively.

Statistically, in a multicarrier CVQKD setting, the following relation can be written between a single-carrier $x_{U_k,j}$ and subcarrier $x_{U_k,i}$ Gaussian quadrature component:

$$x_{U_k,i} = \sum_{j=-\infty}^{\infty} x_{U_k,j} e^{\mathrm{i} j \theta_{\varphi_{U_k,j}}} \, , \qquad (31)$$

and

$$x_{U_k,j} = \frac{1}{2\pi} \int_{-\pi}^{\pi} x_{U_k,i} e^{-\mathrm{i} t_j \theta_{\varphi_{U_k,j}}} \, d\theta_{\varphi_{U_k,j}} \, . \qquad (32)$$

Specifically, for any $\sigma_\omega^2 < \sigma_{\omega_0}^2$, it follows that

$$\Omega \neq 1 \qquad (33)$$

and

$$\left| \theta_{\varphi_{U_k,j}} \right| < \pi \, , \qquad (34)$$

therefore $x_{U_k,i}$ in (31) can be rewritten as

$$x_{U_k,i} = \frac{1}{\Omega} \sum_{j=-\infty}^{\infty} x_{U_k,j} e^{\mathrm{i} j \theta_{\varphi_{U_k,j}} \frac{1}{\Omega}}. \qquad (35)$$

Note that in (31) it is assumed that the integral of (32) exists and is invertible, thus $x_{U_k,j}$ is either square-integrable or absolutely integrable [28].

The $x'_{U_k,i}$ noisy version of (31) is available for Bob via a corresponding $M$ measurement operator (e.g., homodyne or heterodyne measurement) performed on the noisy $\phi'_i$ CV state, as

$$\begin{aligned} x'_{U_k,i} &= M\big(\phi'_i\big) \\ &= M\big(\mathcal{N}_{U_k,i}\big(\phi_i\big)\big). \end{aligned} \qquad (36)$$

In particular, the $\mathcal{S}\left(e^{-\mathrm{i}\theta_{\varphi_{U_k,j}}}\right)$ spectral density of $x_{U_k,j}$ can be defined via the $\left| x'_{U_k,i} \right|^2$ expectation value of $x'_{U_k,i}$, as

$$\mathcal{S}\left(e^{-\mathrm{i}\theta_{\varphi_{U_k,j}}}\right) = \mathbb{E}\left(\left| x'_{U_k,i} \right|^2\right), \qquad (37)$$

which is a statistical measure of the strength of the fluctuations of the subcarrier components [2], [28].

It can be verified that (37) is analogous to the power spectrum $\mathcal{P}\left(e^{-\mathrm{i}\theta_{\varphi_{U_k,j}}}\right)$ of $x_{U_k,j}$,

$$\mathcal{S}\left(e^{-\mathrm{i}\theta_{\varphi_{U_k,j}}}\right) = \mathcal{P}\left(e^{-\mathrm{i}\theta_{\varphi_{U_k,j}}}\right), \qquad (38)$$

where $\mathcal{P}\left(e^{-i\theta_{\varphi_{U_{k},j}}}\right) \geq 0$ is a real function of $\theta_{\varphi_{U_{k},j}}$,

$$\mathcal{P}\left(e^{-i\theta_{\varphi_{U_{k},j}}}\right) = \sum_{g=-\infty}^{\infty} \mathcal{A}_{x_{U_{k},j}}\left(g\right) e^{-i\theta_{\varphi_{U_{k},j}}g}, \qquad (39)$$

such that

$$\mathcal{P}\left(e^{-i\theta_{\varphi_{U_{k},j}}}\right) = \mathcal{P}\left(e^{i\theta_{\varphi_{U_{k},j}}}\right), \qquad (40)$$

where $\mathcal{A}_{x_{U_{k},j}}\left(\cdot\right)$ is the autocorrelation function (autocorrelation sequence [25-28]) of $x_{U_{k},j}$, expressed as

$$\mathcal{A}_{x_{U_{k},j}}\left(g\right) = \mathbb{E}\left(x_{U_{k},j+g} x_{U_{k},j}\right). \qquad (41)$$

Without loss of generality, (37) and (39), allow us to write

$$\mathcal{S}\left(e^{i\theta_{\varphi_{U_{k},j}}}\right) = \mathbb{E}\left(\left|x'_{U_{k},i}\right|^2\right). \qquad (42)$$

Using (42), the estimation of $U^{-1}\left(x_{U_{k},j}\right)$, where $U^{-1}\left(\cdot\right)$ is the inverse CVQFT unitary operation, is expressed as

$$\begin{aligned}
E\left(U^{-1}\left(x_{U_{k},j}\right)\right) &= \mathcal{S}\left(e^{i\theta_{\varphi_{U_{k},j}}}\right) \\
&= \mathcal{P}\left(e^{i\theta_{\varphi_{U_{k},j}}}\right),
\end{aligned} \qquad (43)$$

which, by using (39) can be further evaluated as

$$\begin{aligned}
E\left(U^{-1}\left(x_{U_{k},j}\right)\right) &= \sum_{g=-\infty}^{\infty} \mathcal{A}_{x_{U_{k},j}}\left(g\right) e^{-i\theta_{\varphi_{U_{k},j}}g} \\
&= \mathbb{E}\left(\left|x'_{U_{k},i}\right|^2\right).
\end{aligned} \qquad (44)$$

In particular, $E\left(U^{-1}\left(x_{U_{k},j}\right)\right)$ allows us to uniquely specify $\mathcal{A}_{x'_{U_{k},i}}\left(g\right)$ of a noisy subcarrier quadrature $x'_{U_{k},i}$ as follows.

For a noisy subcarrier quadrature $x'_{U_{k},i}$ of the $i$-th subcarrier CV $\phi'_{U_{k},i}$ of $U_{k}$,

$$\mathcal{A}_{x'_{U_{k},i}}\left(g\right) = \mathcal{A}_{x_{U_{k},j}}\left(\Omega g\right), \qquad (45)$$

where $\Omega$ is defined in (30), while $\mathcal{A}_{x_{U_{k},j}}\left(g\right)$ of $x_{U_{k},j}$ is as

$$\begin{aligned}
\mathcal{A}_{x_{U_{k},j}}\left(g\right) &= \tfrac{1}{2\pi} \int_{-\pi}^{\pi} E\left(U^{-1}\left(x_{U_{k},j}\right)\right) \mathcal{G}_{i}\left(e^{i\theta_{\varphi_{U_{k},j}}}\right) e^{ig\theta_{\varphi_{U_{k},j}}} d\theta_{\varphi_{U_{k},j}} \\
&= \tfrac{1}{2\pi} \int_{-\pi}^{\pi} \mathcal{S}\left(e^{-i\theta_{\varphi_{U_{k},j}}}\right) \mathcal{G}_{i}\left(e^{i\theta_{\varphi_{U_{k},j}}}\right) e^{ig\theta_{\varphi_{U_{k},j}}} d\theta_{\varphi_{U_{k},j}} \\
&= \tfrac{1}{2\pi} \int_{-\pi}^{\pi} \mathcal{P}\left(e^{i\theta_{\varphi_{U_{k},j}}}\right) \mathcal{G}_{i}\left(e^{i\theta_{\varphi_{U_{k},j}}}\right) e^{ig\theta_{\varphi_{U_{k},j}}} d\theta_{\varphi_{U_{k},j}},
\end{aligned} \qquad (46)$$

where $\mathcal{G}_i \left( e^{\mathrm{i}\theta_{\varphi_{U_k,j}}} \right)$ is defined as

$$\mathcal{G}_i \left( e^{\mathrm{i}\theta_{\varphi_{U_k,j}}} \right) = \mathcal{T}_i \left( e^{\mathrm{i}\theta_{\varphi_{U_k,j}}} \right) \mathcal{T}_i \left( \frac{1}{e^{\mathrm{i}\theta_{\varphi_{U_k,j}}}} \right), \tag{47}$$

where

$$\mathcal{T}_i \left( e^{\mathrm{i}\theta_{\varphi_{U_k,j}}} \right) = \begin{cases} T_i \left( \mathcal{N}_{U_k,i} \right), \text{ if } \left| \theta_{\varphi_{U_k,j}} \right| \le \frac{\pi}{\Omega}, \\ 0, \text{ otherwise.} \end{cases} \tag{48}$$

Note that $T_i \left( \mathcal{N}_{U_k,i} \right)$ can be determined via a pilot CV state-based channel estimation procedure; for details, see [8].

The $H \left( x_{U_k,j} \right)$ entropy rate of the Gaussian quadrature component $x_{U_k,j}$ of $\varphi_{U_k,j}$ is evaluated as

$$H \left( x_{U_k,j} \right) = \tfrac{1}{2} \ln 2\pi + \tfrac{1}{2} + \tfrac{1}{4\pi} \int\limits_{-\pi}^{\pi} \ln \mathcal{P} \left( e^{-\mathrm{i}\theta_{\varphi_{U_k,j}}} \right) d\theta_{\varphi_{U_k,j}}. \tag{49}$$

The GQI scheme for multicarrier CVQKD is briefly summarized in Fig. 2. For further details see [10].

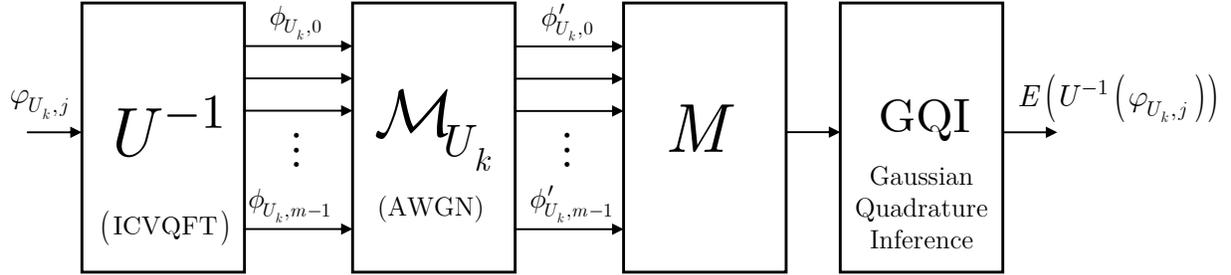

**Figure 2.** The Gaussian Quadrature Inference (GQI) for multicarrier CVQKD. User $U_k$ is equipped with a logical channel $\mathcal{M}_{U_k} = \left[ \mathcal{N}_{U_k,0}, ..., \mathcal{N}_{U_k,m-1} \right]^T$, which has $m$ sub-channels. The input of $U_k$ is $\varphi_{U_k,j} = x_{U_k,j} + \mathrm{i}p_{U_k,j}$, which is transformed via the $U^{-1}$ ICVQFT operation. The $E \left( U^{-1} \left( \varphi_{U_k,j} \right) \right)$ estimate of the $j$-th input CV of $U_k$, $\varphi_{U_k,j} = x_{U_k,j} + \mathrm{i}p_{U_k,j}$ is yielded from the $m$ noisy Gaussian subcarrier CVs, $\phi'_{U_k,i} = x'_{U_k,i} + \mathrm{i}p'_{U_k,i}, i = 0, ..., m-1$. (ICVQFT – inverse continuous-variable quantum Fourier transform, operation $M$ refers to a heterodyne or homodyne measurement, respectively).

Note that for a Gaussian WSS $x \left( n \right)$,

$$H \left( x \right) = H \left( \mathcal{P}_x \right), \tag{50}$$

since $H \left( \cdot \right)$ is a functional of $\mathcal{P}_x \left( \cdot \right)$ [25-28], and

$$H \left( \mathcal{S}_x \right) = H \left( x \right) = H \left( \mathcal{P}_x \right). \tag{51}$$

# 3  Quadrature Evolution by GQI

**Theorem 1** (Quadrature evolution by Gaussian quadrature inference). *Let* $\varphi_{U_k,j} = x_{U_k,j} + \mathrm{i} p_{U_k,j}$ *be the $j$-th input CV of user $U_k$, $k = 0, \ldots, K-1$, let $E\left(U^{-1}\left(\varphi_{U_k,j}\right)\right)$ be the output of GQI, and let $\kappa_{U_k,j} = F\left(M\left(\phi'_{U_k,i}\right)\right)$, $i = 0, \ldots, m-1$ be the F-transformed $m$ noisy subcarrier CVs of $U_k$ at a measurement $M$. The $E\left(\varphi_{U_k,j}\right)$ estimate of $\varphi_{U_k,j}$ is $E\left(\varphi_{U_k,j}\right) = \xi_j \kappa_{U_k,j}$, where $\xi_j$ is an optimal least-squares estimator yielded from $E\left(U^{-1}\left(\varphi_{U_k,j}\right)\right)$.*

*Proof.*
Let

$$\vec{\varphi}_{U_k} = \left(\varphi_{U_k,0}, \ldots, \varphi_{U_k,d-1}\right)^T \tag{52}$$

be a $d$-dimensional input of user $U_k$, $k = 0, \ldots, K-1$, where $\varphi_{U_k,j}$ refers to the $j$-th single-carrier CV, $0 \leq j \leq d-1$, where $\varphi_{U_k,j} = x_{U_k,j} + \mathrm{i} p_{U_k,j}$, $\left\{x_{U_k,j}, p_{U_k,j}\right\}$ are Gaussian random quadratures.

For simplicity, in later parts we refer only to the quadrature component $x_{U_k,j}$ of $\varphi_{U_k,j}$, which formulates the $d$-dimensional vector $\vec{x}_{U_k}$ as

$$\vec{x}_{U_k} = \left(x_{U_k,0}, \ldots, x_{U_k,d-1}\right)^T. \tag{53}$$

Let $x'_{U_k,i}$ be the $i$-th subcarrier quadrature component,

$$x'_{U_k,i} = M\left(\phi'_{U_k,i}\right) \tag{54}$$

resulting from a measurement $M$, where $\phi'_{U_k,i}$ is the $i$-th noisy Gaussian subcarrier CV of $U_k$.
Let

$$\kappa_{U_k,j} = F\left(x'_{U_k,i}\right), i = 0, \ldots, m-1, \tag{55}$$

refer to the F-transformed (FFT) $m$ noisy subcarrier CVs of $U_k$ associated to $\varphi_{U_k,j}$.

Applying the $F$-operation for the $x'_{U_k,i}$, $0 \leq i \leq m-1$, subcarriers of each $j$ of $\vec{x}_{U_k}$ yields the $d$-dimensional vector

$$\vec{\kappa} = \left(\kappa_{U_k,0}, \ldots, \kappa_{U_k,d-1}\right)^T. \tag{56}$$

The QE block uses the GQI output $E\left(U^{-1}\left(x_{U_k,j}\right)\right)$, the $\kappa_{U_k,j}$ elements, and an optimal least-squares estimator $\xi_j$, evaluated via the covariances $\mathrm{cov}_{x_{U_k,j} \kappa_{U_k,j}}, \mathrm{cov}_{\kappa_{U_k,j} \kappa_{U_k,j}}$ as

$$\xi_j = \mathrm{cov}_{x_{U_k,j} \kappa_{U_k,j}} \mathrm{cov}^{-1}_{\kappa_{U_k,j} \kappa_{U_k,j}}, \tag{57}$$

from which $E\left(x_{U_k,j}\right)$ is defined as

$$E\left(x_{U_k,j}\right) = \xi_j \kappa_{U_k,j}.$$
(58)

Let $E\left(x_{U_k,j}\right)$ refer to the estimate of $x_{U_k,j}$, and let error component $\mathfrak{e}_j$ be defined as

$$\mathfrak{e}_j = x_{U_k,j} - E\left(x_{U_k,j}\right),$$
(59)

with an expected error variance $E\left(\sigma^2_{\xi_j}\right)$,

$$E\left(\sigma^2_{\xi_j}\right) = E\left(\mathfrak{e}_j^2\right).$$
(60)

In particular, to determine the $d$-dimensional estimator $\xi$, we introduce the covariance matrices

$$\mathrm{cov}_{\vec{x}_{U_k}\vec{x}_{U_k}}, \mathrm{cov}_{\vec{\kappa}\vec{\kappa}}$$
(61)

and the cross-correlation [29] matrix

$$\mathrm{cov}_{\vec{x}_{U_k}\vec{\kappa}}.$$
(62)

Specifically, based on a $d$-dimensional $\vec{x}_{U_k}$, the covariance matrix $\mathrm{cov}_{\vec{x}_{U_k}\vec{x}_{U_k}}$ is a $d \times d$ dimensional matrix, evaluated as

$$\left[\mathrm{cov}_{\vec{x}_{U_k}\vec{x}_{U_k}}\right]_{qr} = \hat{\mathcal{A}}_{x_{U_k,j}}\left(q - r\right),\ 1 \le q,r \le d,$$
(63)

where $\hat{\mathcal{A}}_{x_{U_k,j}}\left(g\right)$ is the autocorrelation coefficient [29] associated with input $x_{U_k,j}$,

$$\hat{\mathcal{A}}_{x_{U_k,j}}\left(g\right) = \frac{1}{2\pi}\int_{-\pi}^{\pi} E\left(U^{-1}\left(x_{U_k,j}\right)\right) e^{\mathrm{i}g\theta_{\varphi_{U_k,j}}}\,d\theta_{\varphi_{U_k,j}},$$
(64)

where $E\left(U^{-1}\left(x_{U_k,j}\right)\right)$ is the GQI output.

Based on the $d$-dimensional $\vec{\kappa}$, and a GQI output $E\left(U^{-1}\left(x_{U_k,j}\right)\right)$, the $\mathrm{cov}_{\vec{\kappa}\vec{\kappa}}$ covariance matrix is expressed as

$$\left[\mathrm{cov}_{\vec{\kappa}\vec{\kappa}}\right]_{qt} = \mathcal{A}_{\kappa_{U_k,j}}\left(q - t\right),\ 1 \le q,t \le d,$$
(65)

where

$$\mathcal{A}_{\kappa_{U_k,j}}\left(g\right) = \frac{1}{2\pi}\int_{-\pi}^{\pi} E\left(U^{-1}\left(x_{U_k,j}\right)\right)\mathcal{G}_j\left(e^{\mathrm{i}\theta_{\varphi_{U_k,j}}}\right) e^{\mathrm{i}g\theta_{\varphi_{U_k,j}}}\,d\theta_{\varphi_{U_k,j}},$$
(66)

where $\mathcal{G}_j\left(e^{\mathrm{i}\theta_{\varphi_{U_k,j}}}\right)$ is as

$$\mathcal{G}_j\left(e^{\mathrm{i}\theta_{\varphi_{U_k,j}}}\right) = \mathcal{T}_j\left(e^{\mathrm{i}\theta_{\varphi_{U_k,j}}}\right)\mathcal{T}_j\left(\frac{1}{e^{\mathrm{i}\theta_{\varphi_{U_k,j}}}}\right),$$
(67)

and where

$$\mathcal{T}_j\left(e^{\mathrm{i}\theta_{\varphi_{U_k,j}}}\right) = \begin{cases} T_j\left(\mathcal{N}_{U_k,j}\right), & \text{if } \left|\theta_{\varphi_{U_k,j}}\right| \le \frac{\pi}{\Omega}, \\ 0, & \text{otherwise} \end{cases}$$
(68)

such that for a noisy observation $\kappa_{U_k,j}$,

$$\left|T_j\left(\mathcal{N}_{U_k,j}\right)\right| < 1.$$
(69)

Without loss of generality, the cross-correlation matrix $\text{cov}_{\vec{x}_{U_k}\vec{\kappa}}$ is evaluated as

$$\left[\text{cov}_{\vec{x}_{U_k}\vec{\kappa}}\right]_{qt} = \sum_{g=-\infty}^{\infty} \tau_j\left(g\right)\hat{\mathcal{A}}_{x_{U_k,j}}\left(\left(q-1-\Omega\left(t-1\right)\right)-g\right)$$

$$= \sum_{g=-\infty}^{\infty} \tau_j\left(g\right)\hat{\mathcal{A}}_{x_{U_k,j}}\left(\left(\Omega-\Omega t+q-1\right)-g\right), \tag{70}$$

where $\hat{\mathcal{A}}_{x_{U_k,j}}\left(\cdot\right)$ is given in (64), while $\tau_j\left(\cdot\right)$ is determined via (45) as

$$\mathcal{A}_{x_{U_k,j}}\left(g\right) = \left(\tau_j\left(g\right) * \tau_j\left(-g\right)\right) * \hat{\mathcal{A}}_{x_{U_k,j}}\left(g\right), \tag{71}$$

where $*$ is the linear convolution, and

$$\hat{\mathcal{A}}_{x_{U_k,j}}\left(\Omega g\right) = \hat{\mathcal{A}}_{x_{U_k,j}}\left(g\right), \tag{72}$$

while $\hat{\mathcal{A}}_{x_{U_k,i}}\left(\cdot\right)$ refers to the autocorrelation coefficient associated with the $x_{U_k,i}$ subcarrier of user $U_k$.

As follows, (70) can be rewritten as

$$\left[\text{cov}_{\vec{x}_{U_k}\vec{\kappa}}\right]_{qt} = \sum_{g=-\infty}^{\infty} \tau_j\left(g\right)\frac{1}{2\pi}\int_{-\pi}^{\pi} E\left(U^{-1}\left(x_{U_k,j}\right)\right)e^{\text{i}\left(\left(\Omega-\Omega t+q-1\right)-g\right)\theta_{x_{U_k,j}}}d\theta_{\varphi_{U_k,j}}, \tag{73}$$

by fundamental theory [28-29].

Particularly, from (65) and (70), the $\xi$ optimal least-squares estimator for $d$-dimensional vectors is defined via $\text{cov}_{\vec{x}_{U_k}\vec{\kappa}}$ and $\text{cov}_{\vec{\kappa}\vec{\kappa}}$ as

$$\xi = \text{cov}_{\vec{x}_{U_k}\vec{\kappa}}\,\text{cov}_{\vec{\kappa}\vec{\kappa}}^{-1}, \tag{74}$$

from which $E\left(\vec{x}_{U_k}\right)$ is defined precisely as

$$E\left(\vec{x}_{U_k}\right) = \xi\vec{\kappa}\,. \tag{75}$$

Specifically, the estimation $E\left(\vec{x}_{U_k}\right)$ in (75) provides a $d$-dimensional error vector

$$\mathfrak{e} = \vec{x}_{U_k} - E\left(\vec{x}_{U_k}\right), \tag{76}$$

which has a covariance $\text{cov}_{\mathfrak{e}\mathfrak{e}}$ as

$$\text{cov}_{\mathfrak{e}\mathfrak{e}} = \text{cov}_{\vec{x}_{U_k}\vec{x}_{U_k}} - \xi\,\text{cov}_{\vec{x}_{U_k}\vec{\kappa}}^T$$

$$= \text{cov}_{\vec{x}_{U_k}\vec{x}_{U_k}} - \text{cov}_{\vec{x}_{U_k}\vec{\kappa}}\,\text{cov}_{\vec{\kappa}\vec{\kappa}}^{-1}\text{cov}_{\vec{x}_{U_k}\vec{\kappa}}^T, \tag{77}$$

such that the overall expected estimation error, $E\left\{\|\mathfrak{e}\|\right\}$, is minimal,

$$\min E\left\{\|\mathfrak{e}\|^2\right\} = Tr\left(\text{cov}_{\mathfrak{e}\mathfrak{e}}\right), \tag{78}$$

by some fundamental theory [28-29].

The $\mathcal{A}_i\left(g\right)$ autocorrelation coefficients are determined via the inference method of the GQE framework [10], as

$$\mathcal{A}_i\left(g\right) = \frac{1}{2\pi} \int\limits_{-\pi}^{\pi} \left| \frac{\left|\mathcal{T}_i\left(e^{\mathrm{i}\theta_{\varphi_{U_k,j}}}\right)\right|^2 e^{\mathrm{i}\theta_{\varphi_{U_k,j}}g}}{\left|\mathcal{T}_0\left(e^{\mathrm{i}\theta_{\varphi_{U_k,j}}}\right)\right|^2 \lambda_0 \cos\left(\theta_{\varphi_{U_k,j}}\right)} \right| d\theta_{\varphi_{U_k,j}}, \tag{79}$$

where $\lambda_0$ is a Lagrangian coefficient.

∎

The QE quadrature evolution scheme is depicted in Fig. 3.

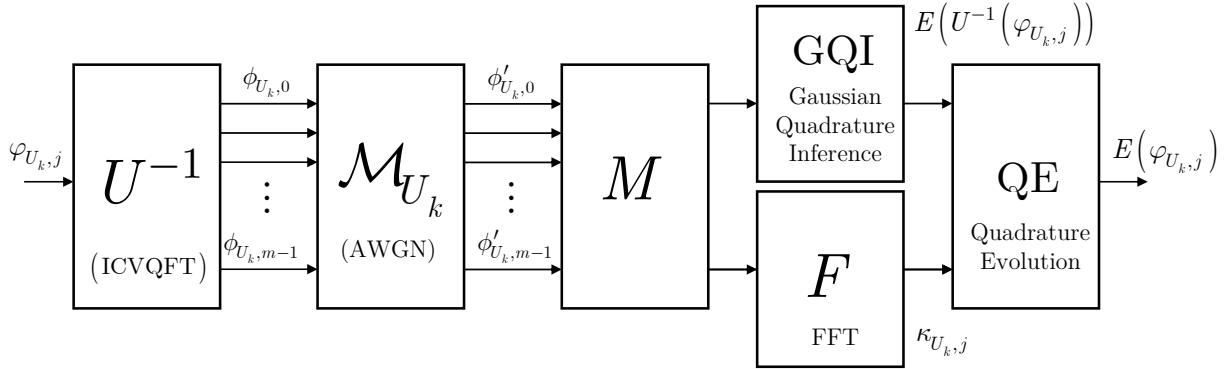

**Figure 3.** Quadrature evolution (QE) by Gaussian quadrature inference (GQI). The Gaussian subcarrier CVs that are output from $\mathcal{M}_{U_k}$ of $U_k$ are measured by $M$, and fed into the GQI module. The measurement result is also transformed by the $F$ (fast Fourier transform) operation. The QE module gets inputs $E\left(U^{-1}\left(\varphi_{U_k,j}\right)\right)$ produced by GQI and $\kappa_{U_k,j} = F\left(M\left(\phi'_{U_k,i}\right)\right)$, $i = 0,\dots,m-1$, and outputs the $E\left(\varphi_{U_k,j}\right)$ estimate of $\varphi_{U_k,j}$ with a theoretical error minimum.

## 3.1 Statistical Secret Key Rate from QE

**Lemma 1.** The $S\left(\mathcal{M}_{U_k}^{(m)}\right)$ statistical secret key rate converges to the $P\left(\mathcal{M}_{U_k}\right)$ statistical private classical information of $\mathcal{M}_{U_k}$, as $m \to \infty$.

*Proof.*

To derive the statistical secret key rate from the statistical quadrature evolution, we can directly apply the results to the achievable secret key rates of the GQI block from [10].

Let sub-index $\left(j, Z, m\right)$ refer to the $j$-th optimal single-carrier at $Z$ autocorrelation coefficients and $m$ sub-channels in $\mathcal{M}_{U_k}$ dedicated to $U_k$. Recalling the results of Theorem 3 and Lemma 1 from [10], the following relation holds at $m$ and $m+1$ sub-channels in $\mathcal{M}_{U_k}$:

$$\begin{aligned} D_{AB}\left(\mathcal{P}_{\hat{x}'_{(j;Z,m)},U_k}\left(e^{\mathrm{i}\theta_{\varphi_{U_k,j}}}\right) \middle\| \mathcal{P}_{\hat{x}_{j},U_k}\left(e^{\mathrm{i}\theta_{\varphi_{U_k,j}}}\right)\right) &\leq D_{AB}\left(\mathcal{P}_{\hat{x}'_{(j;Z,m+1)},U_k}\left(e^{\mathrm{i}\theta_{\varphi_{U_k,j}}}\right) \middle\| \mathcal{P}_{\hat{x}_{j},U_k}\left(e^{\mathrm{i}\theta_{\varphi_{U_k,j}}}\right)\right) \\ &\leq D_{AB}\left(\mathcal{P}'_{\hat{x}_{j},U_k}\left(e^{\mathrm{i}\theta_{\varphi_{U_k,j}}}\right) \middle\| \mathcal{P}_{\hat{x}_{j},U_k}\left(e^{\mathrm{i}\theta_{\varphi_{U_k,j}}}\right)\right), \end{aligned} \tag{80}$$

where $D_{AB}\left(\cdot\right)$ is the relative entropy between Alice and Bob, while $\mathcal{P}_{\hat{x}'_{(j,Z),U_k}}\left(\cdot\right)$ and $\mathcal{P}_{\hat{x}'_{(j,Z),E}}\left(\cdot\right)$ refer to the spectral densities of Bob and Eve.

Therefore, without loss of generality,

$$H\left(\mathcal{P}_{\hat{x}'_{(j,Z,m),U_k}}\left(e^{\mathrm{i}\theta_{\varphi_{U_{k,j}}}}\right)\right) \geq H\left(\mathcal{P}_{\hat{x}'_{(j,Z,m+1),U_k}}\left(e^{\mathrm{i}\theta_{\varphi_{U_{k,j}}}}\right)\right) \geq H\left(\mathcal{P}_{\hat{x}'_{j,U_k}}\left(e^{\mathrm{i}\theta_{\varphi_{U_{k,j}}}}\right)\right). \tag{81}$$

Let $S\left(\mathcal{M}_{U_k}^{(m)}\right)$ refer to the statistical secret key rate of $U_k$ at $m$ sub-channels in $\mathcal{M}_{U_k}^{(m)}$. For the GQI method, it is proven [10] that at $Z \to \infty$,

$$S\left(\mathcal{M}_{U_k}^{(m)}\right) \leq S\left(\mathcal{M}_{U_k}^{(m+1)}\right), \tag{82}$$

where $\mathcal{M}_{U_k}^{(m)}$ and $\mathcal{M}_{U_k}^{(m+1)}$ refer to the logical channel of $U_k$ at $m$ and $m+1$ sub-channels, and $\hat{x}_{(j,Z,m),U_k}$, $\hat{x}'_{(j,Z,m),U_k}$ are the optimal quadratures of Alice and Bob, respectively.

Therefore, at a reverse reconciliation, for $Z \to \infty$ and $m \to \infty$,

$$\begin{aligned}
S\left(\mathcal{M}_{U_k}^{(m)}\right) &= \lim_{n \to \infty} \frac{1}{n} \lim_{Z \to \infty} \max_{\forall x} \left(D_{AB}\left(\hat{x}'_{(j,Z),U_k} \,\middle\|\, \hat{x}_{(j,Z),U_k}\right) - D_{BE}\left(\hat{x}'_{(j,Z),E} \,\middle\|\, \hat{x}'_{(j,Z),U_k}\right)\right) \\
&= \lim_{n \to \infty} \frac{1}{n} \lim_{Z \to \infty} \max_{\forall x} \frac{1}{4\pi} \Bigg(\int_{-\pi}^{\pi} \left(\mathcal{P}_{\hat{x}'_{(j,Z),U_k}}\left(e^{\mathrm{i}\theta_{\varphi_{U_{k,j}}}}\right) - \ln \mathcal{P}_{\hat{x}'_{(j,Z),U_k}}\left(e^{\mathrm{i}\theta_{\varphi_{U_{k,j}}}}\right) - 1\right) d\theta_{\hat{\varphi}_{U_{k,j}}} \\
&\quad - \left(\int_{-\pi}^{\pi} \left(\mathcal{P}_{\hat{x}'_{(j,Z),E}}\left(e^{\mathrm{i}\theta_{\varphi_{U_{k,j}}}}\right) - \ln \mathcal{P}_{\hat{x}'_{(j,Z),E}}\left(e^{\mathrm{i}\theta_{\varphi_{U_{k,j}}}}\right) - 1\right) d\theta_{\hat{\varphi}_{U_{k,j}}}\right)\Bigg),
\end{aligned} \tag{83}$$

where $D_{BE}\left(\cdot\right)$ is the relative entropy between Bob and Eve.

∎

# 4 Conclusions

We defined the QE method for multicarrier CVQKD. The QE scheme extends the results of the GQI to evolve a minimal error estimate of the continuous input regime from noisy discrete variables and estimates. The QE model achieves a theoretically minimized magnitude error, allowing one to evolve the non-observable, single-carrier, continuous variable Gaussian quadratures from the discrete variables with a vanishing error. The QE output is always a unique and stable solution, which minimizes the overall expected estimation error for each individual component of the input. We proved that in a multicarrier CVQKD setting the QE method achieves a vanishing error as the number of subcarriers increases. We derived the corresponding expected error and expected variance via the statistical model. The QE scheme provides a viable, easily implementable, and computationally efficient way to maximize the extractable information from the observed data. The QE framework can be established in an arbitrary CVQKD protocol and measurement setting and is implementable by standard low-complexity functions, which is particularly convenient for experimental CVQKD scenarios.

# Acknowledgements

This work was partially supported by the GOP-1.1.1-11-2012-0092 (*Secure quantum key distribution between two units on optical fiber network*) project sponsored by the EU and European Structural Fund, by the Hungarian Scientific Research Fund - OTKA K-112125, and by the COST Action MP1006.

# Supplemental Information

## S.1  Numerical Evidence

This section proposes numerical evidence to demonstrate the results of the GQI and QE phases through a multiuser multicarrier CVQKD environment (AMQD-MQA [3]). The numerical evidence serves demonstration purposes.

### S.1.1  Parameters

The numerical evidence focuses on the probability distributions of the CV states and subcarriers and studies the statistical properties and effects of modulation variance. It also revises the noise characteristic of the sub-channels and the expected errors of the quadrature estimation procedure.

The

$$x_{U_k, j} \in \mathbb{N}\left(0, \sigma_{\omega_0}^2\right) \tag{S.1}$$

single-carrier inputs of user $U_k$ have a modulation variance of $\sigma_{\omega_0}^2$ and formulate a $d$-dimensional input vector $\vec{x}_{U_k}$ ((53)).

The $j$-th single-carrier is dedicated to a single-carrier channel $\mathcal{N}_{U_k, j}$. The single-carrier channel transmittance coefficient is depicted by $T\left(\mathcal{N}_{U_k, j}\right)$, $0 \leq j \leq d-1$, where $d$ is the dimension of the input vector.

The single-carriers are granulated into $m$ subcarriers, where the $i$-th subcarrier is

$$x_{U_k, i} \in \mathbb{N}\left(0, \sigma_\omega^2\right), \tag{S.2}$$

and has a modulation variance of $\sigma_\omega^2$.

The $m$ sub-channels, $\mathcal{N}_{U_k, i}$, $0 \leq i \leq m-1$, formulate the $\mathcal{M}_{U_k}$ logical channel of user $U_k$, $\mathcal{M}_{U_k} = \left[\mathcal{N}_{U_k, 0}, \ldots, \mathcal{N}_{U_k, m-1}\right]^T$. The $\Delta_{x_i} \in \mathbb{N}\left(0, \sigma_{\mathcal{N}_{U_k, i}}^2\right)$ noise variable of $\mathcal{N}_{U_k, i}$ is added to the subcarriers, where $\sigma_{\mathcal{N}_{U_k, i}}^2$ is the noise variance of $\mathcal{N}_{U_k, i}$. The transmittance coefficient of $\mathcal{N}_{U_k, i}$ is depicted by $T\left(\mathcal{N}_{U_k, i}\right)$.

The $T\left(\mathcal{N}_{U_k, i}\right)$ sub-channel transmittance coefficients are estimated in a pre-communication phase via the subcarrier spreading technique [8]. The subcarrier spreading is an iterative method that uses $p_x$ pilot CV states (pilot: contains no valuable information) to statistically determine the sub-channel transmittance coefficients with minimal theoretical error.

The outputs of the $\mathcal{N}_{U_k,i}$ sub-channels (noisy Gaussian subcarriers) are referred to as

$$x'_{U_k,i} \in \mathbb{N}\left(0, \sigma^2_{x'_{U_k,i}}\right), \tag{S.3}$$

where

$$\sigma^2_{x'_{U_k,i}} = \sigma^2_\omega + \sigma^2_{\mathcal{N}_{U_k,i}}. \tag{S.4}$$

The output of the GQI block is as

$$E\left(U^{-1}\left(x_{U_k,j}\right)\right) \in \mathbb{N}\left(0, \tilde{\sigma}^2_{x'_{U_k,i}}\right), \tag{S.5}$$

which at a direct-GQI (DGQI [10]) can be rewritten as

$$E\left(U^{-1}\left(x_{U_k,j}\right)\right) = F^{-1}\left(x'_{U_k,j}\right) * F^{-1}\left(\beta_{i,\varepsilon}\right), \tag{S.6}$$

where $*$ is the linear convolution, while function $\beta_{i,\varepsilon}$ provides an $\varepsilon$ minimal magnitude error [10],

$$\varepsilon = \arg\min \varepsilon_{\max}, \tag{S.7}$$

where

$$\varepsilon_{\max} = \max_{\forall x}\left|\left|x_{U_k,j}\right|^2 - \left|x'_{U_k,j}\right|^2\right|, \tag{S.8}$$

and $x_{U_k,j}$, $x'_{U_k,j}$ are the input-output single-carrier quadratures, $F^{-1}\left(\cdot\right)$ is the inverse FFT operation [10], while

$$\beta_{i,\varepsilon} = 1 + \sum_{y=1}^{P} C_y \cos\left(yQ_i\right), \tag{S.9}$$

where $C_0$ is arbitrarily set to unity [29] and $P$ is the number of $C_y$ coefficients, while

$$Q_i = \frac{2\pi i}{m}, i = 0, ..., m-1. \tag{S.10}$$

As follows, using (S.6), $\tilde{\sigma}^2_{x'_{U_k,i}}$ in (S.5) is evaluated as

$$\begin{aligned}
\tilde{\sigma}^2_{x'_{U_k,i}} &= \left(F^{-1}\left(\beta_{i,\varepsilon}\right)\right)^2 \sigma^2_{x'_{U_k,i}} \\
&= \left(F^{-1}\left(\beta_{i,\varepsilon}\right)\right)^2 \left(\sigma^2_\omega + \sigma^2_{\mathcal{N}_{U_k,i}}\right),
\end{aligned} \tag{S.11}$$

where

$$F^{-1}\left(\beta_{i,\varepsilon}\right) = \sum_{i=0}^{m-1} \beta_{i,\varepsilon} e^{\frac{i2\pi ij}{m}}, j = 0, ..., n-1. \tag{S.12}$$

The $F$-transformed subcarriers formulate a $d$-dimensional vector $\vec{\kappa}$ ((56)), with $j$-th element as

$$\kappa_{U_k,j} \in \mathbb{N}\left(0, \sigma^2_{\kappa_{U_k,j}}\right), \tag{S.13}$$

and with variance

$$\sigma^2_{\kappa_{U_k,j}} = \sigma^2_{\omega_0} + \sigma^2_{\mathcal{N}_{U_k,j}}. \tag{S.14}$$

The output of the QE block is a $d$-dimensional vector, with $j$-th element

$$E\left(x_{U_k,j}\right) \in \mathbb{N}\left(0,\tilde{\sigma}^2_{\omega_0}\right),\tag{S.15}$$

with variance

$$\tilde{\sigma}^2_{\omega_0} = \beta^2_{i,\varepsilon}\left(\sigma^2_{\omega_0} + \sigma^2_{\mathcal{N}_{U_k,j}}\right),\tag{S.16}$$

where $\beta_{i,\varepsilon}$ is given in (S.9).

The $j$-th single-carrier is estimated from $E\left(U^{-1}\left(x_{U_k,j}\right)\right)$ and $\kappa_{U_k,j}$ as given in (58).

The $\mathfrak{e}$ estimation error is a $d$-dimensional vector with $j$-th element $\mathfrak{e}_j$, see (59). The $E\left(\sigma^2_{\xi_j}\right)$ expected error variance is evaluated via (60).

## S.1.2 Modulation, sub-channel noise

This analysis focuses on the probability distributions, the statistical properties of the input CV states and subcarriers, and the noise characteristics.

In Fig. S.1(a), the Gaussian single-carrier inputs $x_{U_k,j} \in \mathbb{N}\left(0,\sigma^2_{\omega_0}\right)$, $\sigma^2_{\omega_0} = 225$, of user $U_k$ for a $d = 1000$ dimensional input vector $\vec{x}_{U_k}$ are depicted. In Fig. S.1(b), the quadrature component noise $\Delta_{x_i} \in \mathbb{N}\left(0,\sigma^2_{\mathcal{N}_{U_k,i}}\right)$ of the $i$-th Gaussian sub-channel $\mathcal{N}_{U_k,i}$ formulating the logical channel $\mathcal{M}_{U_k}$ of $U_k$ is depicted, $m = 20$, $\sigma^2_{\mathcal{N}_{U_k,i}} = 16$. The data unit index refers to single-carrier and subcarrier CV units, respectively.

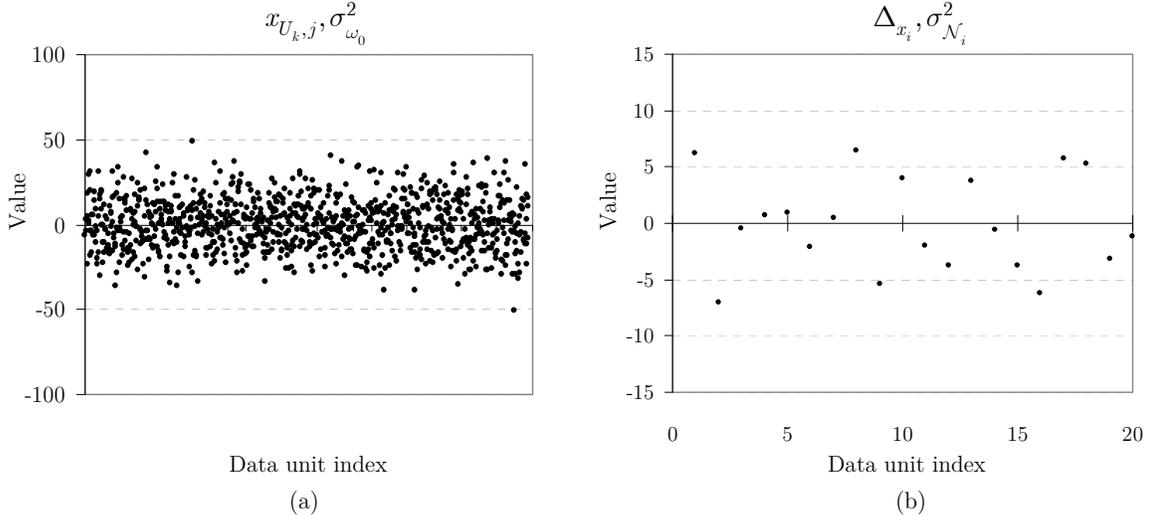

(a)

(b)

**Figure S.1.** (a) Single-carrier inputs, $x_{U_k,j} \in \mathbb{N}\left(0,\sigma^2_{\omega_0}\right)$, $\sigma^2_{\omega_0} = 225$, $0 \le j \le d-1$, $d = 1000$. (b) The $\Delta_{x_i} \in \mathbb{N}\left(0,\sigma^2_{\mathcal{N}_{U_k,i}}\right)$ noise of the $m$ sub-channels of $U_k$, $\sigma^2_{\mathcal{N}_{U_k,i}} = 16$, $0 \le i \le m-1$, $m = 20$.

In Fig. S.2, the $x'_{U_k,i} \in \mathbb{N}\left(0, \sigma^2_{x'_{U_k,i}}\right)$, $\sigma^2_{x'_{U_k,i}} = \sigma^2_\omega + \sigma^2_{\mathcal{N}_{U_k,i}}$, outputs of the $\mathcal{N}_{U_k,i}$ sub-channels are depicted, $\sigma^2_\omega = 64, \sigma^2_{\mathcal{N}_{U_k,i}} = 16, m = 20$.

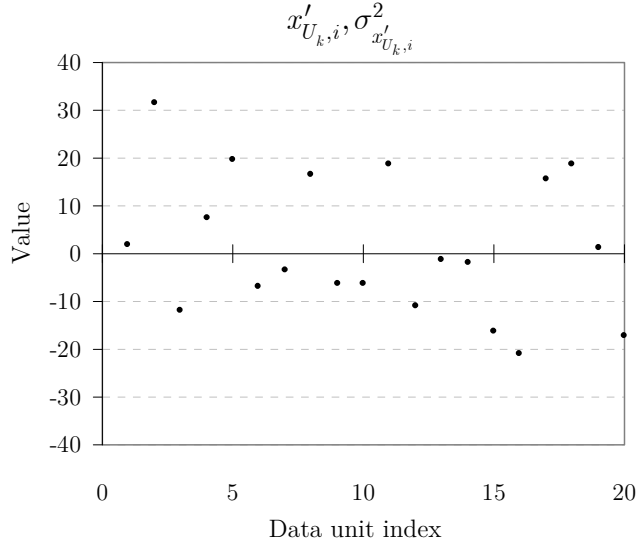

**Figure S.2.** Noisy Gaussian subcarriers, $x'_{U_k,i} \in \mathbb{N}\left(0, \sigma^2_{x'_{U_k,i}}\right)$, $\sigma^2_{x'_{U_k,i}} = \sigma^2_\omega + \sigma^2_{\mathcal{N}_{U_k,i}}$, $\sigma^2_\omega = 64, \sigma^2_{\mathcal{N}_{U_k,i}} = 16, \ 0 \leq i \leq m-1, \ m = 20$.

In the next phase, the GQI block operates on the $x'_{U_k,i}$, $0 \leq i \leq m-1$, $m$ noisy subcarriers to determine $E\left(U^{-1}\left(x_{U_k,j}\right)\right)$ with minimal theoretical error.

## S.1.3 GQI method

This analysis studies the distribution statistics of the GQI output and the FFT-transformed subcarrier elements.

The $E\left(U^{-1}\left(x_{U_k,j}\right)\right) \in \mathbb{N}\left(0, \tilde{\sigma}^2_{x'_{U_k,i}}\right)$, $\tilde{\sigma}^2_{x'_{U_k,i}} = \left(F^{-1}\left(\beta_{i,\varepsilon}\right)\right)^2 \sigma^2_{x'_{U_k,i}}$, output of the GQI block is shown in Fig. S.3(a). The $i$-th subcarrier is estimated via $F^{-1}\left(x'_{U_k,j}\right) * F^{-1}\left(\beta_{i,\varepsilon}\right)$.

In Fig. S.3(b), the $\kappa_{U_k,j}$ elements are depicted, $\kappa_{U_k,j} \in \mathbb{N}\left(0, \sigma^2_{\kappa_{U_k,j}}\right)$, $\sigma^2_{\kappa_{U_k,j}} = \sigma^2_{\omega_0} + \sigma^2_{\mathcal{N}_{U_k,j}}$.

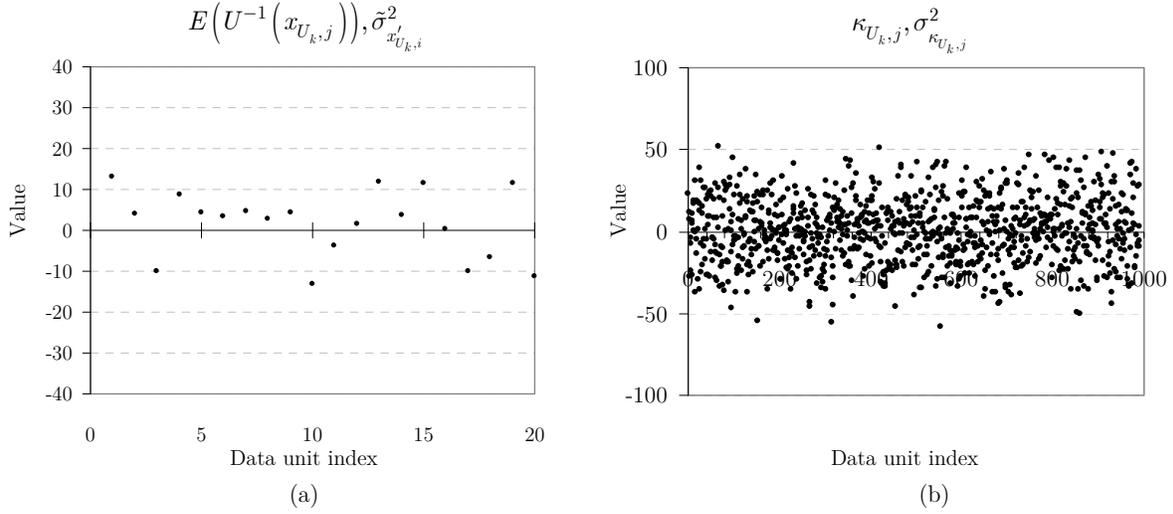

(a)

(b)

**Figure S.3.** (a) The output of GQI, $E\left(U^{-1}\left(x_{U_k,j}\right)\right) \in \mathbb{N}\left(0, \tilde{\sigma}^2_{x'_{U_k,i}}\right),\ 0 \leq i \leq m-1,\ m = 20$. (b) The $\kappa_{U_k,j} \in \mathbb{N}\left(0, \sigma^2_{\kappa_{U_k,j}}\right)$ elements, $0 \leq j \leq d-1,\ d = 1000$.

The GQI output elements provide vanishing magnitude error [10], which will be further utilized in the QE block, since the QE block operates on $E\left(U^{-1}\left(x_{U_k,j}\right)\right)$ and $\kappa_{U_k,j},\ 0 \leq j \leq d-1$, to determine $E\left(x_{U_k,j}\right) \in \mathbb{N}\left(0, \tilde{\sigma}^2_{\omega_0}\right)$.

### S.1.4  QE method

This analysis reveals the distribution statistics of the QE output and the expected estimation error and expected error variance.

In Fig. S.4(a), the output of the QE block, $E\left(x_{U_k,j}\right) \in \mathbb{N}\left(0, \tilde{\sigma}^2_{\omega_0}\right),\ \tilde{\sigma}^2_{\omega_0} = \beta^2_{i,\varepsilon}\left(\sigma^2_{\omega_0} + \sigma^2_{\mathcal{N}_{U_k,j}}\right)$, where $\beta_{i,\varepsilon}$ is given in (S.9), $d = 1000$. The $j$-th single-carrier is estimated as $E\left(x_{U_k,j}\right) = \mathrm{cov}_{x_{U_k,j}\kappa_{U_k,j}} \mathrm{cov}^{-1}_{\kappa_{U_k,j}\kappa_{U_k,j}} \kappa_{U_k,j}$ via $E\left(U^{-1}\left(x_{U_k,j}\right)\right)$ and $\kappa_{U_k,j}$.

In Fig. S.4(b), the elements of the estimation error vector are depicted for $d = 1000$, $\mathfrak{e}_j = x_{U_k,j} - E\left(x_{U_k,j}\right)$.

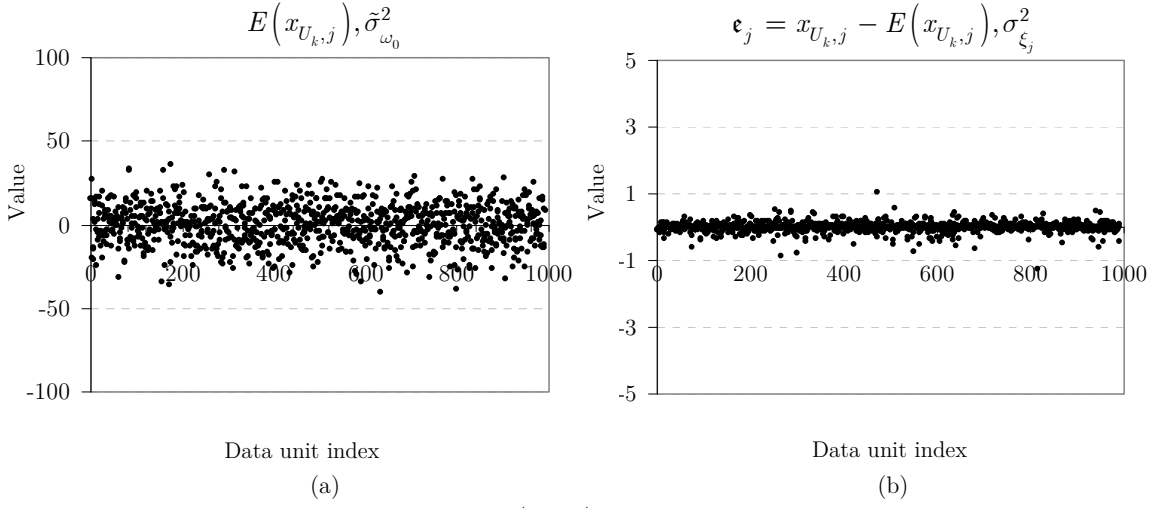

(a)

(b)

**Figure S.4.** (**a**) The output of QE, $E\left(x_{U_{k,j}}\right) = \mathrm{cov}_{x_{U_{k,j}}\kappa_{U_{k,j}}} \mathrm{cov}^{-1}_{\kappa_{U_{k,j}}\kappa_{U_{k,j}}} \kappa_{U_{k,j}}$, $0 \le j \le d-1$, $d = 1000$. (**b**) The components of error vector $\mathfrak{e}$, $\mathfrak{e}_j = x_{U_{k,j}} - E\left(x_{U_{k,j}}\right)$, $0 \le j \le d-1$, $d = 1000$.

The $\mathrm{cov}_{x_{U_{k,j}}\kappa_{U_{k,j}}} \mathrm{cov}^{-1}_{\kappa_{U_{k,j}}\kappa_{U_{k,j}}} \kappa_{U_{k,j}}$ estimate approximates the $j$-th input element, $x_{U_{k,j}}$, with vanishing error, and the $\mathfrak{e}_j$ components converge to zero.

In Fig. S.5, the $E\left(\sigma^2_{\xi_j}\right)$ expected error variances of the noise vector elements $\mathfrak{e}_j$ derived from $\mathrm{cov}_{\mathfrak{ee}}$ are depicted, $d = 1000$.

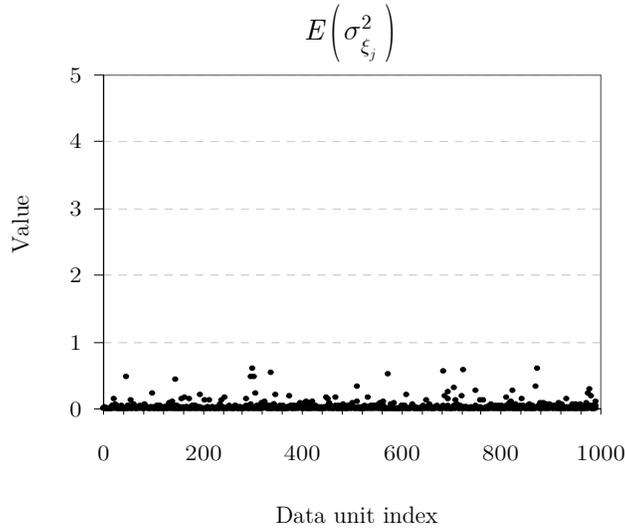

**Figure S.5.** The $E\left(\sigma^2_{\xi_j}\right)$ expected error variances, $0 \le j \le d-1$, $d = 1000$.

### S.1.5 QE expected error and sub-channels

This analysis reveals the expected error variance and errors of the QE method for a $d$-dimensional single-carrier input, in function of the number of sub-channels, $m$. For a given $m$, a $d = 40$ dimensional input is considered.

In Fig. S.6(a), a given $\mathfrak{e}_j$ in function of $m$ is depicted. In Fig. S.6(b), the $E\left(\sigma_{\xi_j}^2\right)$ expected error variance is depicted, at $20 \leq m \leq 45$, $\sigma_{\mathcal{N}_{U_{k,l}}}^2 = 16$, $\sigma_{\omega_0}^2 = 225$ and $d = 40$.

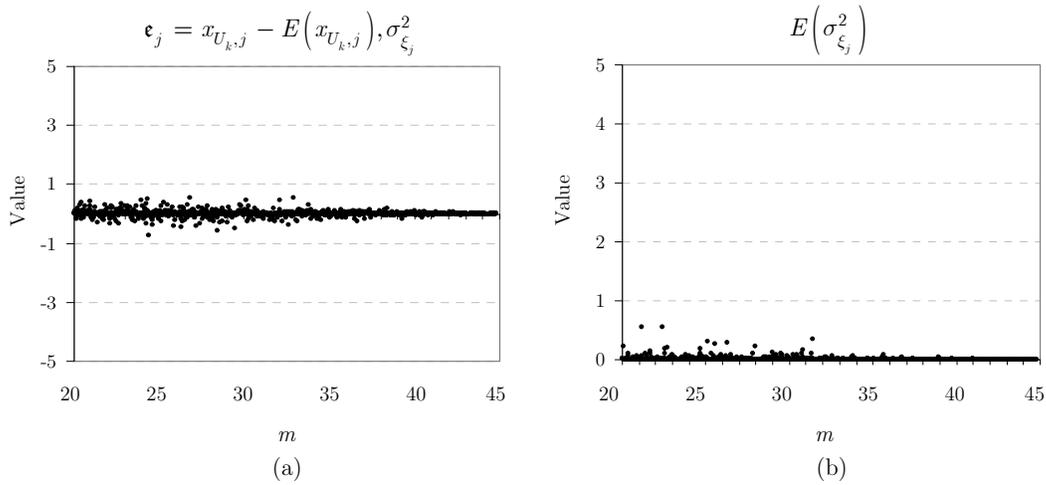

(a)                    (b)

**Figure S.6.** (a) The $\mathfrak{e}_j$ expected error and (b) $E\left(\sigma_{\xi_j}^2\right)$ expected error variance in function of $m$, $20 \leq m \leq 45$, $\sigma_{\mathcal{N}_{U_{k,l}}}^2 = 16$, $\sigma_{\omega_0}^2 = 225$, $d = 40$.

The QE block provides a theoretical error minimum for the quadrature estimation, based on the GQE method. As $m$ increases, $E\left(\sigma_{\xi_j}^2\right)$ converges to zero.

## S.2 Notations

The notations of the manuscript are summarized in Table S.1.

**Table S.1.** Summary of notations.

| | |
|---|---|
| $i$ | Index for the $i$-th subcarrier Gaussian CV, $\left|\phi_i\right\rangle = x_i + \mathrm{i}p_i$. |
| $j$ | Index for the $j$-th Gaussian single-carrier CV, $\left|\varphi_j\right\rangle = x_j + \mathrm{i}p_j$. |
| $l$ | Number of Gaussian sub-channels $\mathcal{N}_i$ for the transmission |
| | of the Gaussian subcarriers. The overall number of the sub-channels is $n$. The remaining $n - l$ sub-channels do not transmit valuable information. |
| $x_i, p_i$ | Position and momentum quadratures of the $i$-th Gaussian subcarrier, $\left\vert \phi_i \right\rangle = x_i + \mathrm{i} p_i$. |
| $x_i', p_i'$ | Noisy position and momentum quadratures of Bob's $i$-th noisy subcarrier Gaussian CV, $\left\vert \phi_i' \right\rangle = x_i' + \mathrm{i} p_i'$. |
| $x_j, p_j$ | Position and momentum quadratures of the $j$-th Gaussian single-carrier $\left\vert \varphi_j \right\rangle = x_j + \mathrm{i} p_j$. |
| $x_j', p_j'$ | Noisy position and momentum quadratures of Bob's $j$-th recovered single-carrier Gaussian CV $\left\vert \varphi_j' \right\rangle = x_j' + \mathrm{i} p_j'$. |
| $x_{A,i}, p_{A,i}$ | Alice's quadratures in the transmission of the $i$-th subcarrier. |
| $\left\vert \phi_i \right\rangle, \left\vert \phi_i' \right\rangle$ | Transmitted and received Gaussian subcarriers. |
| $\mathbf{z} \in \mathcal{CN}\left( 0, \mathbf{K_z} \right)$ | A $d$-dimensional input CV vector to transmit valuable information. |
| $\mathbf{z}'^T$ | A $d$-dimensional noisy output vector, $\mathbf{z}'^T = \mathbf{A}^\dagger \mathbf{z} + \left( F^d\left( \Delta \right) \right)^T = \left( z_0', ..., z_{d-1}' \right)$, where $z_j' = \left( \frac{1}{l} \sum_{i=0}^{l-1} F\left( T_{j,i}\left( \mathcal{N}_{j,i} \right) \right) \right) z_j + F\left( \Delta \right) \in \mathcal{CN}\left( 0, 2\left( \sigma_{\omega_0}^2 + \sigma_{\mathcal{N}}^2 \right) \right)$. |
| $x\left( n \right)$ | A WSS (wide-sense stationary) process. |
| $\mathcal{A}_{x(n)}\left( \cdot \right)$ | Autocorrelation function (sequence) of $x\left( n \right)$. |
| $Z$ | Number of autocorrelation coefficients. |
| $\mathbf{C}_{xx}$ | An $n \times n$ covariance matrix associated with $x\left( n \right)$. |
| $\mathcal{P}_x\left( e^{\mathrm{i}\omega} \right)$ | Power spectrum of $x\left( n \right)$, evaluated as $\mathcal{P}_x\left( e^{\mathrm{i}\omega} \right) = \sum_{l=-\infty}^{+\infty} \mathcal{A}_{x(n)}\left( l \right) e^{-\mathrm{i}\omega l}$, where $\omega \in \left[ -\pi, \pi \right]$. |
| $\mathcal{S}_x\left( e^{\mathrm{i}\omega} \right)$ | Spectral density of $x\left( n \right)$. |
| $H\left( x \right)$ | Entropy rate of $x\left( n \right)$. |
| $f_E\left( x \right)$ | Empirical density function. |
| $D_X(\cdot)$ | Classical relative entropy function. |
| $D(\cdot \| \cdot)$ | Quantum relative entropy function, $$D(\rho \| \sigma) = Tr(\rho \log(\rho)) - Tr(\rho \log(\sigma))$$ $$= Tr[\rho(\log(\rho) - \log(\sigma))],$$ where $\rho$ and $\sigma$ are density matrices. |
| $S(\rho)$ | Neumann entropy, $S(\rho) = -Tr(\rho \log(\rho))$. |
| $N_n(f_E(x))$ | Number of $n$-tuples $(x_1,...,x_n) \in X^n$ with a given empirical density $f_E(x)$. |
| $\mathfrak{L}^p(a,b)$ | Space of Lebesgue-measurable functions. |
| $\mathcal{H}_f(a,b)$ | Functional Hilbert space, $a = -\pi, b = \pi$, $$\mathcal{H}_f(-\pi,\pi) = \mathfrak{L}^2(-\pi,\pi),$$ with a norm $$\|f\|_2 = \tfrac{1}{2\pi}\sqrt{\int_{-\pi}^{\pi}|f(x)|^2\,dx}.$$ |
| $\mathcal{M}_{U_k}$ | Logical channel of user $U_k, k = 0,...,K-1$, where $K$ is the number of total users, $$\mathcal{M}_{U_k} = \left[\mathcal{N}_{U_k,0},...,\mathcal{N}_{U_k,m-1}\right]^T,$$ and $\mathcal{N}_{U_k,i}$ is the $i$-th sub-channel of $\mathcal{M}_{U_k}$, $m$ is the number of subcarriers dedicated to $U_k$. |
| $E\left(U^{-1}\left(\varphi_{U_k,j}\right)\right)$ | Estimate of $U^{-1}\left(\varphi_{U_k,j}\right)$ where $E(\cdot)$ stands for the estimator function, $U^{-1}$ is the inverse CVQFT operation. |
| $E\left(U^{-1}\left(x_{U_k,j}\right)\right)$ | Estimate of $U^{-1}\left(x_{U_k,j}\right)$, where $x_{U_k,j}$ is the quadrature component of $\varphi_{U_k,j}$, where $E(\cdot)$ stands for the estimator function, $U^{-1}$ is the inverse CVQFT operation. |
| $M$ | Measurement operator, homodyne or heterodyne measurement. |
| $\theta_{\varphi_{U_k,j}}$ | $\theta_{\varphi_{U_k,j}} = \pi/\Omega$, where $\Omega = \sigma_{\omega_0}^2/\sigma_\omega^2$, and $\sigma_{\omega_0}^2$, $\sigma_\omega^2$ are the single-carrier and multicarrier modulation variances, $\sigma_\omega^2 < \sigma_{\omega_0}^2$, $\Omega \geq 1$, $\theta_{\varphi_{U_k,j}} \leq \pi$. |
| $\lambda$ | Lagrange multiplier, $$\lambda = \left\| F\left(T^*_{\mathcal{N}}\right)\right\|^2 = \frac{1}{l}\sum_{i=0}^{l-1}\left\| F\left(T^*_i\left(\mathcal{N}_i\right)\right)\right\|^2$$ $$= \frac{1}{l}\sum_{i=0}^{l-1}\left\| \sum_{k=0}^{l-1} T^*_k e^{\frac{-i2\pi ik}{n}}\right\|^2,$$ where $T^*_{\mathcal{N}}$ is the expected transmittance of the $l$ sub-channels under an optimal Gaussian attack. |
| $\tilde{\lambda}_i$ | Optimal Lagrange multipliers. |
| $\mathcal{G}_i\left(e^{i\theta_{\varphi_{U_k,j}}}\right)$ | $$\mathcal{G}_i\left(x\right) = \mathcal{T}_i\left(x\right)\mathcal{T}_i\left(x^{-1}\right), \text{ where}$$ $$\mathcal{T}_i\left(e^{i\theta_{\varphi_{U_k,j}}}\right) = \begin{cases} T_i\left(\mathcal{N}_{U_k,i}\right), \text{ if } \left\|\theta_{\varphi_{U_k,j}}\right\| \leq \frac{\pi}{\Omega}, \\ 0, \text{ otherwise} \end{cases}$$ where $\mathcal{N}_{U_k,i}$ is the $i$-th Gaussian sub-channel of $\mathcal{M}_{U_k}$ of user $U_k$. |
| $\wp$ | Set, defined as $$\wp = \Big\{\frac{1}{2\pi}\int_{-\pi}^{\pi}\mathcal{P}\left(e^{i\theta_{\varphi_{U_k,j}}}\right)\mathcal{G}_i\left(e^{i\theta_{\varphi_{U_k,j}}}\right)e^{i\Omega\theta_{\varphi_{U_k,j}}q}d\theta_{\varphi_{U_k,j}},$$ $$\forall x'_{U_k,i} \in \mathrm{X}_{\phi'_{U_k,i}}, q = 0,...,L-1$$ $$\bigcup \mathcal{P}\left(e^{i\theta_{\varphi_{U_k,j}}}\right) \in \mathfrak{L}^1\left(-\pi,\pi\right),$$ $$\bigcup \mathcal{P}\left(e^{i\theta_{\varphi_{U_k,j}}}\right) \geq 0\Big\}.$$ |
| $\mathrm{X}_{\phi'_{U_k,i}}$ | Set of the autocorrelation functions, $$\mathrm{X}_{\phi'_{U_k,i}} = \left\{\mathcal{A}_{x'_{U_k,0}},...,\mathcal{A}_{x'_{U_k,m-1}}\right\}.$$ |
| $\mathcal{F}_i\left(x\right)$ | Transfer function $$\mathcal{F}_i\left(x\right) = \sum_{q=-\left(L-1\right)}^{L-1} 2\lambda_{iq}x^{-q},$$ where $\lambda_{iq}$ are the Lagrange multipliers, $i = 0,...,m-1, q = 0,...,L-1$. |
| $\Theta$ | Constraint, $\Theta\left(\Gamma\right) = \sum_{i=0}^{m-1}\left(\omega_i\left(\Gamma\right) - \mathcal{A}_{x'_{U_k,i}}\right)^2$. |
| $\Gamma$ | Lagrangian set, $\Gamma = \left[\lambda_0,...,\lambda_{(m-1)}\right]^T$. |
| $F$ | Fourier transform (FFT). |
| $\kappa_{U_k,j}$ | The $F$-transformed $m$ noisy subcarrier CVs of $U_k$, $\kappa_{U_k,j} = F\left(M\left(\phi'_{U_k,i}\right)\right)$, $i = 0,...,m-1$, where $M$ is a measurement operator. |
| $\vec{\varphi}_{U_k}$ | A $d$-dimensional input vector of user $U_k$, $k = 0,...,K-1$, $\vec{\varphi}_{U_k} = \left(\varphi_{U_k,0},...,\varphi_{U_k,d-1}\right)^T$, where $\varphi_{U_k,j}$ refers to the $j$-th single-carrier CV, $\varphi_{U_k,j} = x_{U_k,j} + \mathrm{i}p_{U_k,j}$, and $\left\{x_{U_k,j}, p_{U_k,j}\right\}$ are Gaussian random quadratures. |
| $\vec{x}_{U_k}$ | A $d$-dimensional vector of $U_k$, $\vec{x}_{U_k} = \left(x_{U_k,0},...,x_{U_k,d-1}\right)^T$, where $x_{U_k,j}$ is the quadrature component of $\varphi_{U_k,j}$. |
| $x'_{U_k,i}$ | Noisy subcarrier, discrete variable, $x'_{U_k,i} = M\left(\phi'_{U_k,i}\right)$, where $M$ is a measurement operator, and $\phi'_{U_k,i}$ is the $i$-th noisy Gaussian subcarrier CV of $U_k$. |
| $\vec{\kappa}$ | A $d$-dimensional vector, $\vec{\kappa} = \left(\kappa_{U_k,0},...,\kappa_{U_k,d-1}\right)^T$, resulted from the $F$-operation applied on the subcarriers for each $j$ of $\vec{x}_{U_k}$. |
| $\mathfrak{e}_j$ | Error component, $\mathfrak{e}_j = x_{U_k,j} - E\left(x_{U_k,j}\right)$. |
| $E\left(U^{-1}\left(x_{U_k,j}\right)\right)$ | GQI output. |
| $E\left(x_{U_k,j}\right)$ | QE output, $E\left(x_{U_k,j}\right) = \xi_j \kappa_{U_k,j}$, derived from the GQI output $E\left(U^{-1}\left(x_{U_k,j}\right)\right)$, and $\kappa_{U_k,j}$ via an optimal estimator $\xi_j$. |
| $\mathrm{cov}_{\vec{x}_{U_k}\vec{x}_{U_k}}, \mathrm{cov}_{\vec{\kappa}\vec{\kappa}}$ | Covariance matrices. |
| $\mathrm{cov}_{\vec{x}_{U_k}\vec{\kappa}}$ | Cross-correlation matrix. |
| $\xi$ | Optimal least-squares estimator defined for $d$-dimensional vectors via $\mathrm{cov}_{\vec{x}_{U_k}\vec{\kappa}}$ and $\mathrm{cov}_{\vec{\kappa}\vec{\kappa}}$ as $$\xi = \mathrm{cov}_{\vec{x}_{U_k}\vec{\kappa}}\, \mathrm{cov}_{\vec{\kappa}\vec{\kappa}}^{-1}.$$ |
| $\xi_j$ | Optimal least-squares estimator defined for a $j$-th single-carrier element as $$\xi_j = \mathrm{cov}_{x_{U_k,j}\kappa_{U_k,j}}\, \mathrm{cov}_{\kappa_{U_k,j}\kappa_{U_k,j}}^{-1}.$$ |
| $E\left(x_{U_k,j}\right)$ | Estimate of the $j$-th element via $\xi_j$ as $$E\left(x_{U_{k,j}}\right) = \xi_j \kappa_{U_k,j}$$ $$= \mathrm{cov}_{x_{U_{k,j}}\kappa_{U_{k,j}}} \mathrm{cov}^{-1}_{\kappa_{U_{k,j}}\kappa_{U_{k,j}}} \kappa_{U_{k,j}}.$$ |
| $E\left(\vec{x}_{U_k}\right)$ | A $d$-dimensional QE output, $E\left(\vec{x}_{U_k}\right) = \xi\vec{\kappa}$, where $\xi = \mathrm{cov}_{\vec{x}_{U_k}\vec{\kappa}} \mathrm{cov}^{-1}_{\vec{\kappa}\vec{\kappa}}$. |
| $\mathbf{e}$ | A $d$-dimensional error vector, $\mathbf{e} = \vec{x}_{U_k} - E\left(\vec{x}_{U_k}\right)$. |
| $\mathrm{cov}_{\mathbf{ee}}$ | Covariance of error vector $\mathbf{e}$, $$\mathrm{cov}_{\mathbf{ee}} = \mathrm{cov}_{\vec{x}_{U_k}\vec{x}_{U_k}} - \xi\, \mathrm{cov}^T_{\vec{x}_{U_k}\vec{\kappa}}$$ $$= \mathrm{cov}_{\vec{x}_{U_k}\vec{x}_{U_k}} - \mathrm{cov}_{\vec{x}_{U_k}\vec{\kappa}} \mathrm{cov}^{-1}_{\vec{\kappa}\vec{\kappa}} \mathrm{cov}^T_{\vec{x}_{U_k}\vec{\kappa}}.$$ |
| $E\left(\sigma_\xi^2\right)$ | Expected error variance of $\mathbf{e}$. |
| $E\left(\sigma_{\xi_j}^2\right)$ | Expected error variance of a $j$-th element, $E\left(\sigma_{\xi_j}^2\right) = \mathbf{e}_j^2$. |
| $E\left\{\|\mathbf{e}\|\right\}$ | Overall expected estimation error, $\min E\left\{\|\mathbf{e}\|^2\right\} = Tr\left(\mathrm{cov}_{\mathbf{ee}}\right)$. |
| $\sigma^2_{x'_{U_k,i}}$ | Variance of $x'_{U_k,i}$, $x'_{U_k,i} \in \mathbb{N}\left(0,\sigma^2_{x'_{U_k,i}}\right)$, $\sigma^2_{x'_{U_k,i}} = \sigma^2_\omega + \sigma^2_{\mathcal{N}_{U_k,i}}$. |
| $E\left(U^{-1}\left(x_{U_k,j}\right)\right)$ | GQI output, $E\left(U^{-1}\left(x_{U_k,j}\right)\right) \in \mathbb{N}\left(0,\tilde{\sigma}^2_{x'_{U_k,i}}\right)$, evaluated at a direct-GQI (DGQI [10]) as $$E\left(U^{-1}\left(x_{U_k,j}\right)\right) = F^{-1}\left(x'_{U_k,j}\right) * F^{-1}\left(\beta_{i,\varepsilon}\right),$$ where $*$ is the linear convolution, while function $\beta_{i,\varepsilon}$ provides an $\varepsilon$ minimal magnitude error $$\varepsilon = \arg\min \varepsilon_{\max},$$ where $$\varepsilon_{\max} = \max_{\forall x} \left|\left|x_{U_k,j}\right|^2 - \left|x'_{U_k,j}\right|^2\right|,$$ and $x_{U_k,j}$, $x'_{U_k,j}$ are the input, output single-carrier quadratures, $F^{-1}\left(\cdot\right)$ is the inverse FFT operation [10], while $$\beta_{i,\varepsilon} = 1 + \sum_{y=1}^{P} C_y \cos\left(yQ_i\right),$$ |

| | |
|---|---|
| | where $C_0$ is arbitrarily set to unity [29], $P$ is the number of $C_y$ coefficients, while $$Q_i = \frac{2\pi i}{m}, i = 0,\dots,m-1.$$ |
| $\sigma^2_{\kappa_{U_k,j}}$ | Variance of $\kappa_{U_k,j} \in \mathbb{N}\left(0, \sigma^2_{\kappa_{U_k,j}}\right)$, $\sigma^2_{\kappa_{U_k,j}} = \sigma^2_{\omega_0} + \sigma^2_{\mathcal{N}_{U_k,j}}$. |
| $\tilde{\sigma}^2_{\omega_0}$ | Variance of QE output, $E\left(x_{U_k,j}\right) \in \mathbb{N}\left(0, \tilde{\sigma}^2_{\omega_0}\right)$, evaluated as $$\tilde{\sigma}^2_{\omega_0} = \beta^2_{i,\varepsilon}\left(\sigma^2_{\omega_0} + \sigma^2_{\mathcal{N}_{U_k,j}}\right).$$ |
| $*$ | Linear convolution operator. |
| $\mathfrak{f}$ | Function, provides an $\varepsilon$ minimal magnitude error, for $E\left(U^{-1}\left(x_{U_k,j}\right)\right) = F^{-1}\left(x'_{U_k,j}\right) * F^{-1}\left(\beta_i\right), i = 0,\dots,m-1.$ |
| $\varepsilon$ | Minimal magnitude error, $\varepsilon = \arg\min \varepsilon_{\max}$, where $\varepsilon_{\max} = \max\limits_{\forall x}\left|\left|x_{U_k,j}\right|^2 - \left|x'_{U_k,j}\right|^2\right|$, $x_{U_k,j}$, $x'_{U_k,j}$ are the input, output single-carrier quadratures, $F^{-1}\left(\cdot\right)$ is the inverse FFT operation. |
| $\mathbf{x}_{U_k,i}$ | Input subcarrier vector, $\mathbf{x}_{U_k,i} = \left(x_{U_k,0},\dots,x_{U_k,m-1}\right)^T$. |
| $\mathbf{x}'_{U_k,i}$ | Output subcarrier vector, $\mathbf{x}'_{U_k,i} = \left(x'_{U_k,0},\dots,x'_{U_k,m-1}\right)^T$. |
| $\mathrm{A}_i$ | Operator, $\mathrm{A}_i = 2\sigma^2_\omega \frac{i}{m}$, $i = 0,\dots,\frac{m}{2}$, where $\sigma^2_\omega$ is the subcarrier modulation variance, $m$ is the number of subcarriers of $U_k$. |
| $\mathfrak{E}$ | Estimator function, evaluated as $$\mathfrak{E} = \begin{cases} i = 0 : \mathfrak{E}\left(\mathrm{A}_0\right) = \frac{1}{m^2}\left|x'_{U_k,0}\right|^2 \\ i = 1,\dots,\frac{m}{2}-1 : \mathfrak{E}\left(\mathrm{A}_i\right) = \frac{1}{m^2}\left(\left|x'_{U_k,i}\right|^2 + \left|x'_{U_k,m-i}\right|^2\right), \\ i = \frac{m}{2} : \mathfrak{E}\left(\mathrm{A}_{\frac{m}{2}}\right) = \frac{1}{m^2}\left|x'_{U_k,\frac{m}{2}}\right|^2 \end{cases}.$$ |
| $\alpha$, $\alpha_\varepsilon$ | $\alpha = m\sum\limits_{i=0}^{m-1}\beta^2_i$, $\alpha_\varepsilon = m\sum\limits_{z=0}^{m-1}\beta^2_{z,\varepsilon}$. |
| $\beta_i$ | Parameter, defined to evaluate function $\mathfrak{f}\left(s\right)$ as |
| | $$\mathfrak{f}\left(s\right)=\frac{1}{\alpha}\left|\sum_{i=0}^{m-1}\beta_{i}e^{\frac{i2\pi is}{m}}\right|^{2}$$ $$=\frac{1}{\alpha}\left|\int_{-m/2}^{m/2}\cos\left(\frac{2\pi si}{m}\right)\beta\left(i-\frac{m}{2}\right)di\right|^{2}.$$ |
| $\beta_{i,\varepsilon}$ | Function to minimize $\varepsilon$, $\beta_{i,\varepsilon}=1+\sum_{y=1}^{P}C_{y}\cos\left(yQ_{i}\right)$, where $C_{0}$ is arbitrarily set to unity, $P$ is the number of $C_{y}$ coefficients, while $Q_{i}=\frac{2\pi i}{m}, i=0,\dots,m-1$. |
| $P\left(\mathcal{M}_{U_{k}}\right)$ | Statistical private classical information of $U_{k}$. |
| $S\left(\mathcal{M}_{U_{k}}\right)$ | Statistical secret key rate of $U_{k}$. |
| $\hat{x}_{(j:Z),U_{k}}$ , $\hat{x}'_{(j:Z),U_{k}}$ , $\hat{x}'_{(j:Z),E}$ | Optimal quadratures of Alice, Bob and Eve, obtained at $Z$ autocorrelation coefficients. |
| $\mathcal{X}_{AB}$ | Holevo quantity of Bob's output. |
| $\mathcal{X}_{BE}$ | Holevo information leaked to the Eve in a reverse reconciliation. |
| $\rho_{k}^{AB}$ | Bob's optimal output density matrix. |
| $\rho_{k}^{BE}$ | Eve's optimal density matrix, at a reverse reconciliation. |
| $\sigma^{AB}$ | Bob's optimal output average density matrix. |
| $\sigma^{BE}$ | Eve's average density matrix, at a reverse reconciliation. |
| $U_{K_{out}}$ | The unitary CVQFT operation, $U_{K_{out}}=\frac{1}{\sqrt{K_{out}}}e^{\frac{-i2\pi ik}{K_{out}}}$, $i,k=0,\dots,K_{out}-1$, $K_{out}\times K_{out}$ unitary matrix. |
| $U_{K_{in}}$ | The unitary inverse CVQFT operation, $U_{K_{in}}=\frac{1}{\sqrt{K_{in}}}e^{\frac{i2\pi ik}{K_{in}}}$, $i,k=0,\dots,K_{in}-1$, $K_{in}\times K_{in}$ unitary matrix. |
| $z\in\mathcal{CN}\left(0,\sigma_{z}^{2}\right)$ | The variable of a single-carrier Gaussian CV state, $\left|\varphi_{i}\right\rangle\in\mathcal{S}$. Zero-mean, circular symmetric complex Gaussian random variable, $\sigma_{z}^{2}=\mathbb{E}\left[\left|z\right|^{2}\right]=2\sigma_{\omega_{0}}^{2}$, with i.i.d. zero mean, Gaussian random quadrature components $x,p\in\mathbb{N}\left(0,\sigma_{\omega_{0}}^{2}\right)$, where $\sigma_{\omega_{0}}^{2}$ is the variance. |

| | |
|---|---|
| $\Delta \in \mathcal{CN}\left(0,\sigma_\Delta^2\right)$ | The noise variable of the Gaussian channel $\mathcal{N}$, with i.i.d. zero-mean, Gaussian random noise components on the position and momentum quadratures $\Delta_x, \Delta_p \in \mathbb{N}\left(0,\sigma_\mathcal{N}^2\right)$, $\sigma_\Delta^2 = \mathbb{E}\left[\left|\Delta\right|^2\right] = 2\sigma_\mathcal{N}^2$. |
| $d \in \mathcal{CN}\left(0,\sigma_d^2\right)$ | The variable of a Gaussian subcarrier CV state, $\left|\phi_i\right\rangle \in \mathcal{S}$. Zero-mean, circular symmetric Gaussian random variable, $\sigma_d^2 = \mathbb{E}\left[\left|d\right|^2\right] = 2\sigma_\omega^2$, with i.i.d. zero mean, Gaussian random quadrature components $x_d, p_d \in \mathbb{N}\left(0,\sigma_\omega^2\right)$, where $\sigma_\omega^2$ is the (constant) modulation variance of the Gaussian subcarrier CV state. |
| $F^{-1}\left(\cdot\right) = \mathrm{CVQFT}^\dagger\left(\cdot\right)$ | The inverse CVQFT transformation, applied by the encoder, continuous-variable unitary operation. |
| $F\left(\cdot\right) = \mathrm{CVQFT}\left(\cdot\right)$ | The CVQFT transformation, applied by the decoder, continuous-variable unitary operation. |
| $F^{-1}\left(\cdot\right) = \mathrm{IFFT}\left(\cdot\right)$ | Inverse FFT transform, applied by the encoder. |
| $\sigma_{\omega_0}^2$ | Single-carrier modulation variance. |
| $\sigma_\omega^2 = \frac{1}{l}\sum_l \sigma_{\omega_i}^2$ | Multicarrier modulation variance. Average modulation variance of the $l$ Gaussian sub-channels $\mathcal{N}_i$. |
| $\begin{aligned}\left|\phi_i\right\rangle &= \left|\mathrm{IFFT}\left(z_{k,i}\right)\right\rangle \\ &= \left|F^{-1}\left(z_{k,i}\right)\right\rangle = \left|d_i\right\rangle.\end{aligned}$ | The $i$-th Gaussian subcarrier CV of user $U_k$, where IFFT stands for the Inverse Fast Fourier Transform, $\left|\phi_i\right\rangle \in \mathcal{S}$, $d_i \in \mathcal{CN}\left(0,\sigma_{d_i}^2\right)$, $\sigma_{d_i}^2 = \mathbb{E}\left[\left|d_i\right|^2\right]$, $d_i = x_{d_i} + \mathrm{i}p_{d_i}$, $x_{d_i} \in \mathbb{N}\left(0,\sigma_{\omega_F}^2\right)$, $p_{d_i} \in \mathbb{N}\left(0,\sigma_{\omega_F}^2\right)$ are i.i.d. zero-mean Gaussian random quadrature components, and $\sigma_{\omega_F}^2$ is the variance of the Fourier transformed Gaussian state. |
| $\left|\varphi_{k,i}\right\rangle = \mathrm{CVQFT}\left(\left|\phi_i\right\rangle\right)$ | The decoded single-carrier CV of user $U_k$ from the subcarrier CV, expressed as $F\left(\left|d_i\right\rangle\right) = \left|F\left(F^{-1}\left(z_{k,i}\right)\right)\right\rangle = \left|z_{k,i}\right\rangle$. |
| $\mathcal{N}$ | Gaussian quantum channel. |
| $\mathcal{N}_i, i = 0,\ldots,n-1$ | Gaussian sub-channels. |
| $T\left(\mathcal{N}\right)$ | Channel transmittance, normalized complex random vari- |
| | able, $T(\mathcal{N}) = \operatorname{Re} T(\mathcal{N}) + \mathrm{i}\operatorname{Im} T(\mathcal{N}) \in \mathcal{C}$. The real part identifies the position quadrature transmission, the imaginary part identifies the transmittance of the position quadrature. |
| $T_i(\mathcal{N}_i)$ | Transmittance coefficient of Gaussian sub-channel $\mathcal{N}_i$, $T_i(\mathcal{N}_i) = \operatorname{Re}(T_i(\mathcal{N}_i)) + \mathrm{i}\operatorname{Im}(T_i(\mathcal{N}_i)) \in \mathcal{C}$, quantifies the position and momentum quadrature transmission, with (normalized) real and imaginary parts $0 \leq \operatorname{Re} T_i(\mathcal{N}_i) \leq 1/\sqrt{2}$, $0 \leq \operatorname{Im} T_i(\mathcal{N}_i) \leq 1/\sqrt{2}$, where $\operatorname{Re} T_i(\mathcal{N}_i) = \operatorname{Im} T_i(\mathcal{N}_i)$. |
| $T_{Eve}$ | Eve's transmittance, $T_{Eve} = 1 - T(\mathcal{N})$. |
| $T_{Eve,i}$ | Eve's transmittance for the $i$-th subcarrier CV. |
| $\mathbf{z} = \mathbf{x} + \mathrm{i}\mathbf{p} = (z_0,\dots,z_{d-1})^T$ | A $d$-dimensional, zero-mean, circular symmetric complex random Gaussian vector that models $d$ Gaussian CV input states, $\mathcal{CN}(0,\mathbf{K_z})$, $\mathbf{K_z} = \mathbb{E}[\mathbf{zz}^\dagger]$, where $z_j = x_j + \mathrm{i}p_j$, $\mathbf{x} = (x_0,\dots,x_{d-1})^T$, $\mathbf{p} = (p_0,\dots,p_{d-1})^T$, $x_j \in \mathbb{N}(0,\sigma^2_{\omega_0})$, $p_j \in \mathbb{N}(0,\sigma^2_{\omega_0})$ i.i.d. zero-mean Gaussian random variables. |
| $\mathbf{d} = F^{-1}(\mathbf{z})$ | An $l$-dimensional, zero-mean, circular symmetric complex random Gaussian vector, $\mathcal{CN}(0,\mathbf{K_d})$, $\mathbf{K_d} = \mathbb{E}[\mathbf{dd}^\dagger]$, $\mathbf{d} = (d_0,\dots,d_{l-1})^T$, $d_i = x_i + \mathrm{i}p_i$, $x_i, p_i \in \mathbb{N}(0,\sigma^2_{\omega_F})$ are i.i.d. zero-mean Gaussian random variables. The $i$-th component is $d_i \in \mathcal{CN}(0,\sigma^2_{d_i})$, $\sigma^2_{d_i} = \mathbb{E}[|d_i|^2]$. |
| $\mathbf{y}_k \in \mathcal{CN}\left(0, \mathbb{E}[\mathbf{y}_k\mathbf{y}_k^\dagger]\right)$ | A $d$-dimensional zero-mean, circular symmetric complex Gaussian random vector. |
| $y_{k,m}$ | The $m$-th element of the $k$-th user's vector $\mathbf{y}_k$, expressed as $y_{k,m} = \sum_l F(T_i(\mathcal{N}_i))F(d_i) + F(\Delta_i)$. |
| $F(\mathbf{T}(\mathcal{N}))$ | Fourier transform of $\mathbf{T}(\mathcal{N}) = [T_0(\mathcal{N}_0)\dots,T_{l-1}(\mathcal{N}_{l-1})]^T \in \mathcal{C}^l$, the complex transmittance vector. |

| $F(\Delta)$ | Complex vector, expressed as $F(\Delta) = e^{\frac{-F(\Delta)^T \mathbf{\kappa}_{F(\Delta)} F(\Delta)}{2}}$, with covariance matrix $\mathbf{K}_{F(\Delta)} = \mathbb{E}\left[F(\Delta) F(\Delta)^\dagger\right]$. |
|---|---|
| $\mathbf{y}[j]$ | AMQD block, $\mathbf{y}[j] = F(\mathbf{T}(\mathcal{N})) F(\mathbf{d})[j] + F(\Delta)[j]$. |
| $\tau = \|F(\mathbf{d})[j]\|^2$ | An exponentially distributed variable, with density $f(\tau) = \left(1/2\sigma_\omega^{2n}\right) e^{-\tau/2\sigma_\omega^2}$, $\mathbb{E}[\tau] \leq n2\sigma_\omega^2$. |
| $T_{Eve,i}$ | Eve's transmittance on the Gaussian sub-channel $\mathcal{N}_i$, $T_{Eve,i} = \operatorname{Re} T_{Eve,i} + \mathrm{i}\operatorname{Im} T_{Eve,i} \in \mathcal{C}$, $\quad 0 \leq \operatorname{Re} T_{Eve,i} \leq 1/\sqrt{2}$, $0 \leq \operatorname{Im} T_{Eve,i} \leq 1/\sqrt{2}$, $0 \leq |T_{Eve,i}|^2 < 1$. |
| $d_i$ | A $d_i$ subcarrier in an AMQD block. |
| $\nu_{\min}$ | The $\min\{\nu_0,\ldots,\nu_{l-1}\}$ minimum of the $\nu_i$ sub-channel co-efficients, where $\nu_i = \sigma_{\mathcal{N}}^2 \big/ \left|F(T_i(\mathcal{N}_i))\right|^2$ and $\nu_i < \nu_{Eve}$. |
| $\sigma_\omega^2$ | Constant modulation variance, $\sigma_\omega^2 = \nu_{Eve} - \nu_{\min}\mathcal{G}(\delta)_{p(x)}$, $\nu_{Eve} = \frac{1}{\lambda}$, $\quad \lambda = \left|F(T_{\mathcal{N}}^*)\right|^2 = \frac{1}{n}\sum_{i=0}^{n-1}\left|\sum_{k=0}^{n-1} T_k^* e^{\frac{-\mathrm{i}2\pi ik}{n}}\right|^2$ and $T_{\mathcal{N}}^*$ is the expected transmittance of the Gaussian sub-channels under an optimal Gaussian collective attack. |

# S.3 Abbreviations

| | |
|---|---|
| **AMQD** | **Adaptive Multicarrier Quadrature Division** |
| **AWGN** | **Additive White Gaussian Noise** |
| **CV** | **Continuous-Variable** |
| **CVQFT** | **Continuous-Variable Quantum Fourier Transform** |
| **CVQKD** | **Continuous-Variable Quantum Key Distribution** |
| **DGQI** | **Direct Gaussian Quadrature Inference** |
| **DV** | **Discrete Variable** |
| **FFT** | **Fast Fourier Transform** |
| **GQI** | **Gaussian Quadrature Inference** |
| **ICVQFT** | **Inverse CVQFT** |
| **IFFT** | **Inverse Fast Fourier Transform** |
| **MQA** | **Multiuser Quadrature Allocation** |
| **PDF** | **Probability Density Function** |
| **QE** | **Quadrature Estimation** |
| **QFT** | **Quantum Fourier Transform** |
| **QKD** | **Quantum Key Distribution** |
| **SNR** | **Signal to Noise Ratio** |
| **WSS** | **Wide-Sense Stationary** |